\documentclass{svmult}

\usepackage{makeidx}       
\usepackage{graphicx} 
                       
\usepackage{multicol}      
\usepackage{cite}          
                             
\usepackage[bottom]{footmisc}
\usepackage{color}  

\usepackage{mathdesign}
\usepackage[T1]{fontenc}
 \usepackage[british]{babel}
\usepackage[numbers,comma,sectionbib]{natbib}
\usepackage{graphicx}
\usepackage{booktabs}
\usepackage{amsmath,amssymb}
\usepackage{verbatim}

\newcommand\blfootnote[1]{%
  \begingroup
  \renewcommand\thefootnote{}\footnote{#1}%
  \addtocounter{footnote}{-1}%
  \endgroup
}

\newcommand{\kpsii}{\ket{\psi^0}}
\newcommand{\bpsii}{\bra{\psi^0}}
\newcommand{\Fp}{F_{\mathrm{p}}}

\newcommand{\OddBellp}{ \ket{\Phi^{+}}}
\newcommand{\EvenBellp}{ \ket{\Psi^{+}}}

\newcommand{\BraOddBellp}{ \bra{\Phi^{+}}}

\newcommand{\sz}{\sigma_z}

\newcommand{\sx}{\sigma_x}

\newcommand{\epsp}{\epsilon_\mathrm{p}}
\newcommand{\epse}{\epsilon_\mathrm{e}}
\newcommand{\epso}{\epsilon_\mathrm{o}}

\newcommand{\tp}{\tau_\mathrm{P}}

\newcommand{\Vp}{V_\mathrm{P}}
\newcommand{\rhoq}{\rho}
\newcommand{\rhoi}{\rho^{(0)}}

\newcommand{\rmixed}{\rho_{11,10}}
\newcommand{\rodd}{\rho_{01,10}}
\newcommand{\reven}{\rho_{00,11}}

\newcommand{\ebiteff}{\mathrm{ebits/run}}

\newcommand{\fa}{f_\mathrm{A}}
\newcommand{\fb}{f_\mathrm{B}}

\newcommand{\fp}{f_\mathrm{p}}

\newcommand{\Vint}{V_\mathrm{int}}

\newcommand{\ti}{t_\mathrm{i}}
\newcommand{\tf}{t_\mathrm{f}}

\newcommand{\szA}{\sigma_z^\mathrm{A}}
\newcommand{\szB}{\sigma_z^\mathrm{B}}

\newcommand{\Ojoint}{\mathcal{O}}

\newcommand{\nss}{\bar{n}_{\mathrm{ss}}}
\newcommand{\Mp}{M_\mathrm{P}}

\newcommand{\Vthreshp}{V_{\mathrm{th}+}}
\newcommand{\Vthreshm}{V_{\mathrm{th}-}}
\newcommand{\Vthresh}{V_{\mathrm{th}}}
\newcommand{\Vthr}{V_{\mathrm{th}}}
\newcommand{\conc}{\mathcal{C}}
\newcommand{\LN}{E_\mathcal{N}}

\newcommand{\Ratee}{\mathcal{E}}
\newcommand{\rateE}{\mathcal{E}}
\newcommand{\psuccess}{p_{\mathrm{success}}}

\newcommand{\ket}[1]{\left\lvert #1 \right\rangle}
\newcommand{\bra}[1]{\left\langle #1 \right\rvert}
\newcommand{\avg}[1]{\left\langle #1 \right\rangle}

\newcommand{\tr}{\text{Tr}}

\newcommand{\Mb}{M_B}
\newcommand{\Mc}{M_C}

\newcommand{\Fbg}{\mathrm{Fb}_0}
\newcommand{\Fbe}{\mathrm{Fb}_1}
\newcommand{\Pg}{P_{\ket{0}}}
\newcommand{\Pe}{P_{\ket{1}}}
\newcommand{\Pee}{P_{\ket{2}}}
\newcommand{\Gzo}{\Gamma_{10}}
\newcommand{\Goz}{\Gamma_{01}}
\newcommand{\Got}{\Gamma_{21}}
\newcommand{\Gto}{\Gamma_{12}}

\newcommand{\tinit}{\tau_{\mathrm{init}}}
\newcommand{\Perr}{P_{\mathrm{err}}}

\newcommand{\Vps}{V_{\mathrm{ps}}}
\newcommand{\tlat}{\tau_\mathrm{Fb}}

\newcommand{\K}{\mathrm{K}}
\newcommand{\mK}{\mathrm{mK}}

\newcommand{\kHz}{\mathrm{kHz}}
\newcommand{\MHz}{\mathrm{MHz}}
\newcommand{\GHz}{\mathrm{GHz}}

\newcommand{\us}{\mu\mathrm{s}}
\newcommand{\ns}{\mathrm{ns}}

\newcommand{\m}{\mathrm{m}}

\newcommand{\dB}{\mathrm{dB}}

\newcommand{\zotrans}{0\leftrightarrow1}

\newcommand{\QA}{\mathrm{Q}_{\mathrm{A}}}
\newcommand{\QB}{\mathrm{Q}_{\mathrm{B}}}

\newcommand{\Ma}{M_A}

\newcommand{\wrf}{\omega_{\mathrm{RF}}}

\newcommand{\ChiA}{\chi_{\mathrm{A}}}
\newcommand{\ChiB}{\chi_{\mathrm{B}}}

\newcommand{\betaA}{\beta_{\mathrm{A}}}
\newcommand{\betaB}{\beta_{\mathrm{B}}}
\newcommand{\betaBA}{\beta_{\mathrm{BA}}}

\newcommand{\Rmnum}[1]{\expandafter\@slowromancap\romannumeral #1@}

\newcommand{\wc}{\omega_{\mathrm{c}}}
\newcommand{\Vout}{V_{\mathrm{out}}}

\newcommand{\hl}[1]{#1}

\graphicspath{{../Figures/}}

\begin{document}

\title*{Digital Feedback in Superconducting Quantum Circuits}

\author{Diego Rist\`e \and Leonardo DiCarlo}
\institute{QuTech and Kavli Institute of Nanoscience, Delft University of Technology, 2600 GA Delft, The Netherlands.}

\maketitle

\begin{abstract}
This chapter covers the development of feedback control of superconducting qubits using projective measurement and a discrete set of conditional actions, here referred to as digital feedback. We begin with an overview of the applications of digital feedback in quantum computing. We then introduce an implementation of high-fidelity projective measurement of superconducting qubits. This development lays the ground for closed-loop control based on the binary measurement result. A first application of digital feedback control is fast and deterministic qubit reset, allowing the repeated initialization of a qubit more than an order of magnitude faster than its relaxation rate. A second application employs feedback in a multi-qubit setting to convert the generation of entanglement by parity measurement from probabilistic to deterministic, targeting an entangled state with the desired parity every time.
\end{abstract}

\blfootnote{Part of this chapter appeared in Refs.~\cite{Riste12, Riste12b, Riste13b}.}

\section{Digital feedback control in quantum computing}
Moving from proof-of-principle demonstrations of quantum gates and algorithms to fully fledged quantum hardware requires closing the loop between qubit measurement and control.  
There are different categories of quantum feedback control, depending on the type of measurement and feedback law used. For clarity, we first offer a classification of quantum feedback, similarly to that used in classical feedback. Then, we focus on the particular class of discrete-time, digital feedback. 

\subsection{Classification of quantum feedback}
A first distinction is between \hl{continuous-time} and \hl{discrete-time feedback}. In the first case, measurement and control are continuous in time and concurrent. An example is the stabilization of a qubit state using continuous partial measurement, as discussed in Refs.~\cite{Gillett10, Sayrin11, Bushev06, Koch10, Brakhane12}. In discrete-time feedback, instead, the conditional control is applied only after a measurement has been performed and processed. Here, we focus exclusively on discrete-time implementations. This class can be further divided into two categories, analog and digital.  We speak of \hl{analog feedback} when the measurement result assumes a continuum of values and the feedback law is a continuous function of the result. An example is the experiment in Ref.~\citenum{deLange14}, where the feedback controller first integrates the signal produced by a weak measurement and then applies the resulting coherent operation on the qubit. If the measurement has a finite set of possible results, instead, the possible feedback actions are also finite. We refer to this as \hl{digital feedback}. The simplest example is qubit reset (section~\ref{sec:reset}), in which a strong projective measurement collapses the qubit into either the ground or excited state. Here, a $\pi$ rotation brings the qubit to ground. 
Another interesting example  is digital feedback using ancilla-based partial measurement~\cite{Blok13, Groen13}. In this case, the measurement output is discrete, showing that partial measurement is not necessarily associated with analog feedback. In many applications, digital feedback forces determinism into one of the most controversial aspects of quantum mechanics, namely the measurement, whose result is intrinsically probabilistic. Looking at the action of digital feedback as a black box, we expect to see a definite output qubit state for a given input. In an ideal feedback scheme, measurement results and the conditioned operations vary at every run of the protocol, but the overall process is deterministic and the output state is always the same. For example, one can project a two-qubit superposition to a specific Bell state by combining a parity measurement with digital feedback (section~\ref{sec:ebm}). 

\subsection{Protocols using digital feedback}
\label{sec:fb_protocols}
Several quantum information processing (QIP) protocols call for digital feedback. One of the requirements for a quantum computer is efficient qubit initialization~\cite{DiVincenzo00}. Often, the steady state of a qubit does not correspond to a pure computational state $\ket{0}$ or $\ket{1}$, bur rather to a mixture of the two. Therefore, active initialization methods have been used in many QIP architectures. Examples are laser or microwave initialization~\cite{Monroe95, Atature06, Valenzuela06, Manucharyan09} and initialization by control of the qubit relaxation rate~\cite{Reed10b, Mariantoni11}. An alternative method, recently used with NV centers in diamond~\cite{Robledo11} and superconducting qubits (section~\ref{sec:readout}), relies on projective measurement to initialize the qubits into a pure state. However, measurement alone cannot produce the desired state with certainty, since the measurement result is probabilistic. Closing a feedback loop based on this measurement turns the unwanted outcomes into the desired state. 
A qubit register must not only be initialized in a pure state at the beginning of computation, but often also during the computation. For example, performing multiple rounds of error correction is facilitated by resetting ancilla qubits to their ground state after each parity check~\cite{Schindler11}. When using a qubit as a detector (e.g. of charge~\cite{Riste13} or photon parity~\cite{Sun14}), a fast reset can be used to increase the sampling rate without keeping track of past measurement outcomes.  

Similarly, in the multi-qubit setting, digital feedback is key to turning \\measurement-based protocols from probabilistic to deterministic. An example is the generation of entanglement by parity measurement~\cite{Ruskov03}. A parity measurement projects an initial maximal  superposition state into an entangled state with a well-defined parity, i.e., with either even or odd total number of qubit excitations (section~\ref{sec:ebm}).  However, once again, the outcome of the parity measurement is random. When running the protocol open-loop multiple times, the average final state has no specific parity and is unentangled. Only by forcing a definite parity using feedback can one generate a target entangled state deterministically. 

A variation of closed-loop control, named \hl{feedforward}, applies control on qubits different from those measured. Feedforward schemes have already found application in quantum communication, where the main objective is the secure transmission of quantum information over a distance. In quantum teleportation, a measurement on the Bell basis of two qubits projects a third qubit, at any distance, into the state of the first, to within a single-qubit rotation~\cite{Nielsen00}. The measurement result determines which qubit rotation, if any, must be applied to teleport the original state. 
An extension of teleportation is entanglement swapping~\cite{Nielsen00}. This protocol transfers entanglement to two qubits which never interact, and forms the basis for quantum repeaters~\cite{Briegel98}, aiming to distribute entanglement across larger distances than allowed by a lossy communication channel.
Here, measurement and feedback are used in every step to first purify~\cite{Bennett96} and then deterministically transfer entangled pairs to progressively farther nodes. 

In quantum computing, feedforward operations are at the basis of the first schemes devised to protect a qubit state from errors.  
The simplest protocol is the bit-flip code~\cite{Mermin07}, which encodes the quantum state of one qubit into a an entangled state of three, and uses measurement of two-qubit operators (syndromes) in combination with feedback to correct for $\sx$ (bit-flip) errors. Of similar structure is the phase-flip code, which protects against $\sz$ (phase-flip) errors. To protect against errors on any axis, the minimum size of the encoding is five qubits. In \hl{quantum error correction}, projective measurement is more than a tool to detect discrete errors that have   already occurred. In fact, the measurement serves to discretize the set of possible erorrs. Measuring the error syndromes forces one and only one of these errors to happen. This greatly simplifies the feedback step, which is now restricted to a finite set of correcting actions.     

While few-qubit error correction schemes are capable of correcting any single error, they require currently inaccessible measurement and gate fidelities. A more realistic approach is offered by topologically protected circuits such as surface codes~\cite{Fowler12}, where errors as high as $1\%$ are tolerated at the expense of a larger number of physical qubits required~\cite{Wang11}. One cycle in a surface code, aimed at maintaining a logical state encoded in a square lattice of qubits, includes the projective measurements of 4-qubit operators as error syndromes. When an error is detected on a data qubit, the corrective, coherent feedback operation is replaced by a change of sign in the operators for the following syndrome measurements involving that qubit. In other words, errors are kept track of by the classical controller rather than fixed~\cite{Kelly15, Riste15}. Beyond protecting a state from external perturbations, performing fault-tolerant quantum computing will require robustness to gate errors. In surface codes, single- and two-qubit gates on logical qubits are also based on projective measurements and in some cases require digital feedback to apply conditional rotations~\cite{Fowler12b}. 

In addition to the gate model~\cite{DiVincenzo00}, digital feedback is central to the paradigm of \hl{measurement-based quantum computing}~\cite{Briegel09}. In this approach, also called one-way computation, the initial state is an entangled state of a large number of qubits. All logical operations are performed by projective measurements. To make computation deterministic, feedback selects the measurement bases at each computational step, conditional on the measurement results. 

\subsection{Experimental realizations of digital feedback}
Digital feedback has been employed for entanglement swapping with trapped ions~\cite{Riebe08} and for the unconditional teleportation of photonic~\cite{Furusawa98}, ionic~\cite{Barrett04, Riebe04}, and atomic~\cite{Sherson06, Krauter13} qubits. 
In linear optics, feedforward has been used to implement segments of one-way quantum computing~\cite{Tame06, Prevedel07, Chen07, Vallone08, Ukai11, Bell14} and for photon multiplexing~\cite{Vitelli13}. 
In the solid state, the first approach to feedback, of the analog type, was used to stabilize Rabi oscillations of a superconducting qubit~\cite{Vijay12}. Soon after, digital feedback with high-fidelity projective measurement was introduced in the solid state, also using superconducting circuits~\cite{Riste12b, CampagneIbarcq13}.  
Recently, digital feedback has been extended to multi-qubit protocols with superconducting qubits (section~\ref{sec:ebm} and Ref.~\citenum{Steffen13}) and NV centers in diamond~\cite{Pfaff14}.   

\subsection{Concepts in digital feedback}  
\label{sec:fbbasics}
\begin{figure*}
\centering
\includegraphics[width=0.3\columnwidth]{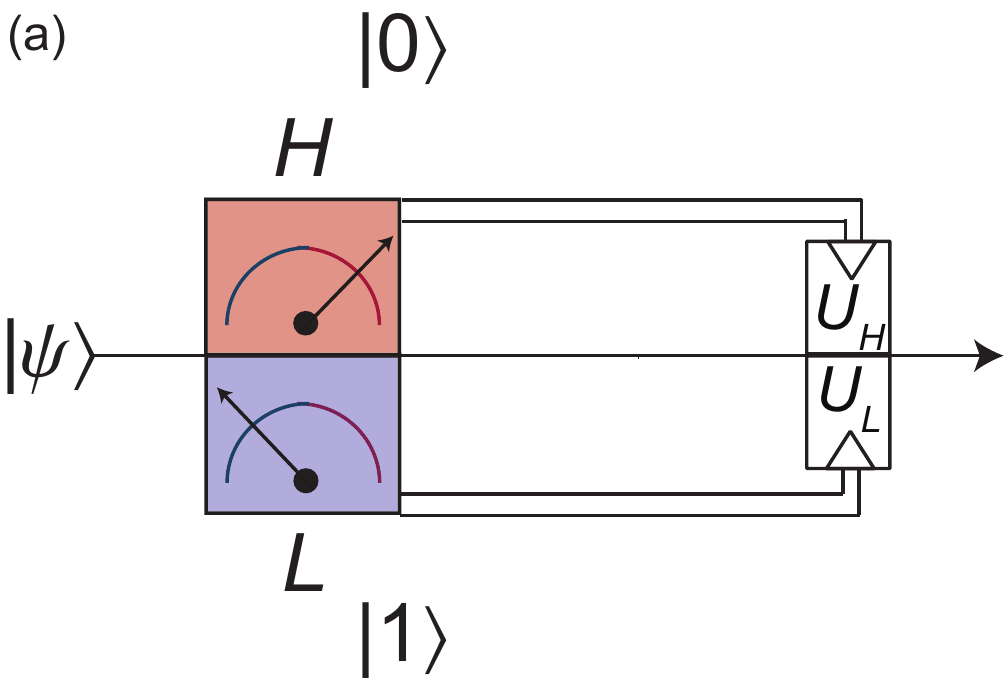} \includegraphics[width=0.3\columnwidth]{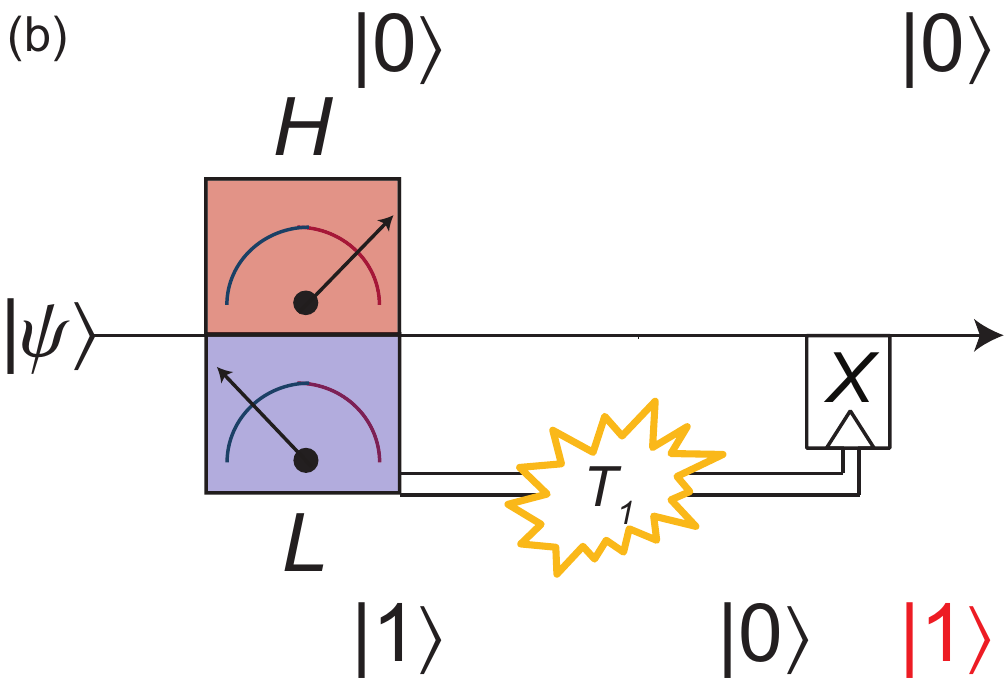} \includegraphics[width=0.3\columnwidth]{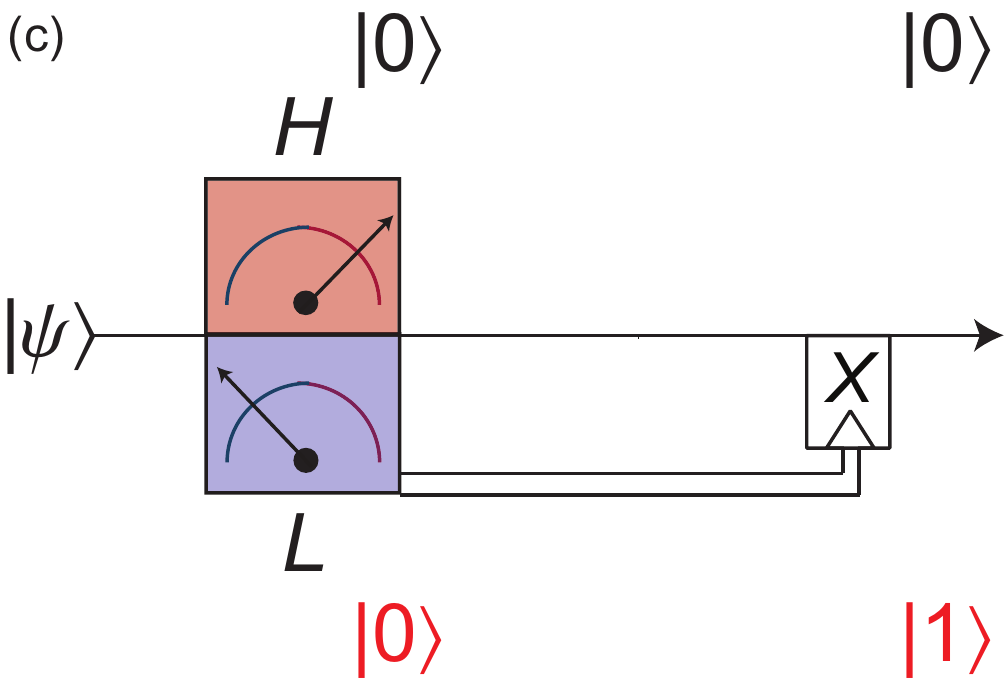}
\caption{\label{fig:1_book}  \textbf{Concept of a single-qubit digital feedback loop and possible errors.} \textbf{a,} The measurement is digitized into either $H$ or $L$ for qubit declared in $\ket{0}$ or $\ket{1}$. A different unitary rotation is applied for each result. Errors occurring in case of qubit relaxation between measurement and action (\textbf{b}) or wrong measurement assignment (\textbf{c}). Top (bottom) row indicates the actual qubit state corresponding to result $H$ ($L$).} 
\end{figure*}

The basic ingredients for a digital feedback loop are: 1) \hl{projective} qubit \hl{readout} and 2) control conditional on the measurement result (see Fig.~\ref{fig:1_book}\textbf{a} for the simplest single-qubit loop). 
The main challenge for (1) is to obtain a high-fidelity readout which is also nondemolition, thus leaving the qubits in a state  consistent with the measurement result. A mismatch between measurement result and post-measurement qubit state will trigger the wrong feedback action (Fig.~\ref{fig:1_book}\textbf{b}). 
The requirement for (2) is to minimize the time, or \hl{latency}, between measurement and conditional action. Various sources contribute to latency: the time for the signal to travel from the sample to the feedback controller, the time for the feedback controller to process the signal and discretize it, and the delay to the execution of the conditional qubit gates. If a transition between levels occurs in one of the measured qubits during this interval, for instance because of spontaneous relaxation, its state becomes inconsistent with the chosen feedback action, resulting in the wrong final state (Fig.~\ref{fig:1_book}\textbf{c}). In feedforward protocols, such as error correction or teleportation, the feedback action is applied to data qubits, which are different from the measured ancilla qubits. In this case, the loop must also be fast compared to the coherence times of data qubits. 

The simplest example of digital feedback is single-qubit \hl{reset}. Here, the qubit is projected by measurement onto $\ket{0}$ or $\ket{1}$ and, depending on the targeted state, a $\pi$ pulse is applied conditional on the measurement result. In this example, we consider the effect of the errors in Fig.~\ref{fig:1_book}~\textbf{b},\textbf{c}, modeling the qubit as a classical three-level system, where the third level includes the possibility of transitions out of the qubit subspace. This is relevant in the case of transmon qubits with a sizeable steady-state excitation~\cite{Riste12b, CampagneIbarcq13}. We indicate with $p^M_{ij}$ the probability of obtaining the measurement result $M$ with initial state $\ket{i}$ and post-measurement state $\ket{j}$. With $\Gamma_{ij}$ we indicate the transition rates from $\ket{i}$ to $\ket{j}$, and with $\tlat$ the time between the end of measurement and the end of the conditional operation.  
For perfect pulses, the combined errors $\Perr^{\theta}$ for initial state $\cos(\theta)\ket{0} + \sin(\theta)\ket{1}$ are, to first order:
\begin{equation}
\label{eq:fb}
\begin{aligned}
&\Perr^{\theta=0} = p^L_{00}+p^H_{01}+\Gamma_{01}\tlat, \\
&\Perr^{\theta=\pi}=  p^H_{11}+p^L_{10}+p_{12}+(\Gamma_{10}+\Gamma_{12})\tlat,
\end{aligned}
\end{equation}
and weighted combinations thereof for other $\theta$. 
A simple way to improve feedback fidelity is to perform two cycles back to back. While the dominant error for $\theta = 0$ remains unchanged, for $\theta = \pi$ it decreases to $\Perr^{\theta=0}+p_{12}+\Gamma_{12}\tlat$. 
The second cycle compensates errors arising from relaxation to $\ket{0}$ between measurement and pulse in the first cycle. However, it does not correct for excitation from $\ket{1}$ to $\ket{2}$. For this reason, adding more cycles does not significantly reduce the error, unless the population in $\ket{2}$ is brought back to the qubit subspace. This can be done~\cite{Riste12b, CampagneIbarcq13} by a deterministic $\pi$ pulse returning the population from $\ket{2}$ to $\ket{1}$, or with a more complex feedback loop capable of resolving and manipulating all three states. 
 
\subsection{Closing the loop in cQED}
\label{sec:fbcqed}
Until recently, the coherence times of superconducting qubits bottlenecked both achievable readout fidelity and required feedback speed. The development of  circuit quantum electrodynamics~\cite{Blais04,Wallraff04} with 3D cavities (3D cQED)~\cite{Paik11} constitutes a watershed. The new order of magnitude in qubit coherence times $(>10~\us$), combined with Josephson parametric amplification~\cite{Castellanos-Beltran08, Vijay09}, allows projective readout with fidelities $~\sim99\%$ and realizing feedback control with off-the-shelf electronics. In the following section, we detail our implementation of high-fidelity projective readout of a transmon qubit in 3D cQED. We then shift focus to the real-time signal processing by the feedback controller, and on the resulting feedback action. 

\begin{figure*}
\centering
\includegraphics[width=0.5\columnwidth]{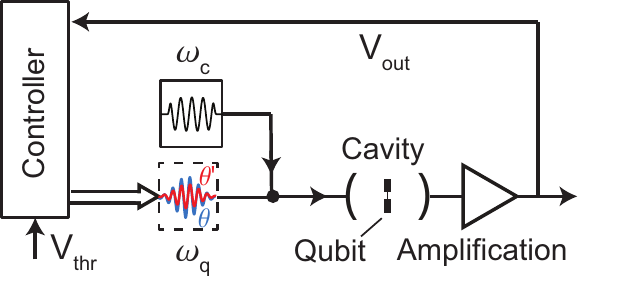}
\caption{\label{fig:fbcqed}  \textbf{Simplified schematic of a single-qubit feedback loop in cQED.} Upon application of a measurement tone at $\wc$, the signal $\Vout$ obtained from processing of the cavity output, carrying information on the qubit state, is input to the feedback controller and compared to a preset threshold $\Vthr$. If $\Vout>\Vthr$ (or $\Vthr<\Vthr$), the conditional rotation $\theta\, (\theta')$ is applied to the qubit.} 
\end{figure*}

\section{High-fidelity projective readout of transmon qubits}
\label{sec:readout}

\subsection{Experimental setup}

Our system consists of an Al 3D cavity enclosing two superconducting transmon qubits, labeled $\QA$ and $\QB$, with transition frequencies $\omega_{\mathrm{A(B)}}/2\pi = 5.606 (5.327)~\GHz$, relaxation times $T_{1\mathrm{A(B)}}=23~(27)~\us$. The fundamental mode of the cavity (TE101) resonates at $\omega_{r}/2\pi=6.548~\GHz$ (for qubits in ground state) with $\kappa/2\pi = 430~\kHz$ linewidth, and couples with $g/2\pi \sim 75~\MHz$ to both qubits. The dispersive shifts~\cite{Wallraff04}
$\chi_\mathrm{A(B)}/\pi=-3.7~(-2.6)~\MHz$, both large compared to $\kappa/2\pi$, place the system in the strong dispersive regime of cQED~\cite{Schuster07}.

\begin{figure}
\center
\includegraphics[width=0.8\columnwidth]{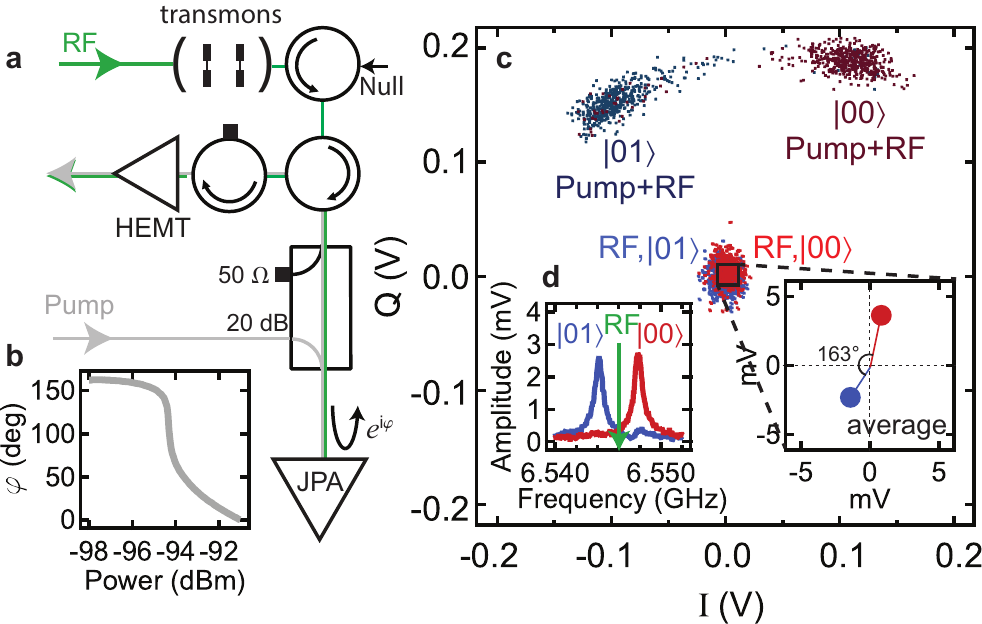}
\caption{\label{fig:1_paper1} \textbf{JPA-backed dispersive transmon readout.} \textbf{a,} Simplified diagram of the experimental setup, showing the input path for the readout signal carrying the information on the qubit state (RF, green) and the stronger, degenerate tone (Pump, grey) biasing the JPA.  Both microwave tones are combined at the JPA and their sum is reflected with a phase dependent on the total power (\textbf{b}), amplifying the small signal. An additional tone (Null) is used to cancel any pump leakage into the cavity. The JPA is operated at the low-signal gain of $\sim25~\dB$ and $2~\MHz$ bandwidth. \textbf{c,} Scatter plot in the $I-Q$ plane for sets of 500 single-shot measurements. Light red and blue: readout signal obtained with an RF tone probing the cavity for qubits in $\ket{00}$ and $\ket{01}$, respectively.
Dark red and blue: the Pump tone is added to the RF. \textbf{d,} Spectroscopy of the cavity fundamental mode for qubits in $\ket{00}$ and $\ket{01}$. The RF frequency is chosen halfway between the two resonance peaks, giving the maximum phase contrast ($163^\circ$, see inset on the right). Figure taken from Ref.~\citenum{Riste12}.}
\end{figure}

Qubit readout in cQED typically exploits dispersive interaction with the cavity. A readout pulse is applied at or near resonance with the cavity, and a coherent state builds up in the cavity with amplitude and phase encoding the multi-qubit state~\cite{Wallraff04,Majer07}. We optimize readout of $\QA$ by injecting a microwave pulse through the cavity at $\wrf=\omega_{r}-\chi_\mathrm{A}$, the average of  the resonance frequencies corresponding to qubits in $\ket{00}$ and $\ket{01}$, with left (right) index denoting the state of $\QB$ ($\QA$) (Figs.~\ref{fig:1_paper1}\textbf{a},\textbf{d}). This choice maximizes the phase difference  between the pointer coherent states. Homodyne detection of the output signal, itself proportional to the intra-cavity state, is overwhelmed by the noise added
by the semiconductor amplifier (HEMT), precluding high-fidelity single-shot readout (Fig.~\ref{fig:1_paper1}\textbf{c}). We introduce a \hl{Josephson parametric amplifier} (JPA)~\cite{Castellanos-Beltran08} at the front end of the amplification chain to boost the readout signal by exploiting the power-dependent phase of reflection at the JPA (see Figs.~\ref{fig:1_paper1}\textbf{a},\textbf{b}). Depending on the qubit state, the weak signal transmitted through the cavity is either added to or subtracted from a much stronger pump tone incident on the JPA, allowing single-shot discrimination between the two cases  (Fig.~\ref{fig:1_paper1}\textbf{c}).

\subsection{Characterization of JPA-backed qubit readout and initialization}

\begin{figure}
\center
\includegraphics[width=0.7\columnwidth]{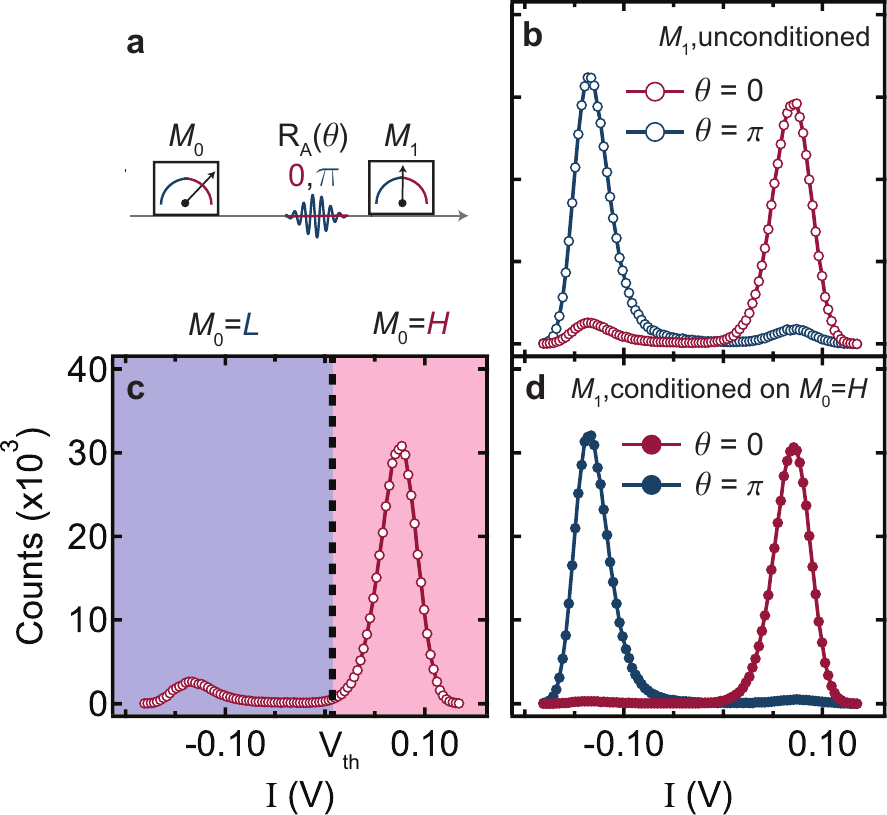}
\caption{\label{fig:2_paper1} \textbf{Ground-state initialization by measurement.} \textbf{a,}  Pulse sequence used to distinguish between the qubit states ($\Mb$), upon conditioning on the result of an initialization measurement $\Ma$. The sequence is repeated every $250~\us$. \textbf{b,}  Histograms of $500\,000$ shots of $\Mb$, without (red) and with (blue) inverting the population of $\QA$ with a $\pi$ pulse. \textbf{c,}  Histograms of $\Ma$, with $V_\mathrm{th}$ indicating the threshold voltage used to digitize the result. \textbf{d,}  $\Mb$ conditioned on $\Ma = H$ to initialize the system in the ground state, suppressing the residual steady-state excitation.
The conditioning threshold, selecting $91\%$ of the shots, matches the value for optimum discrimination of the state of $\QA$. Figure taken from Ref.~\citenum{Riste12}.}
\end{figure}

\begin{figure}
\center
\includegraphics[width=0.9\columnwidth]{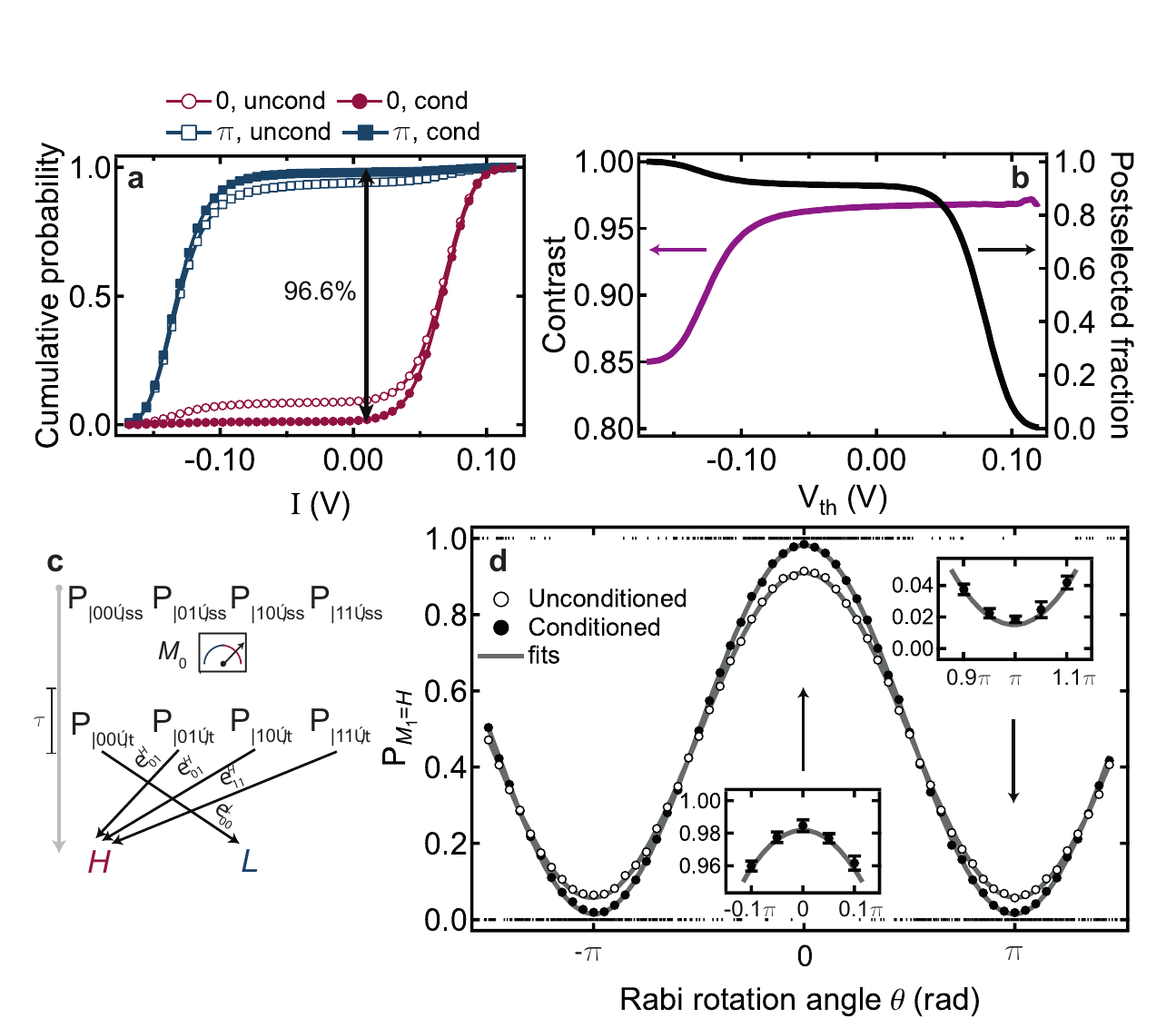}
\caption{\label{fig:3_paper1}\textbf{Analysis of readout fidelity.} \textbf{a,} Cumulative histograms for $\Mb$ without and with conditioning on $\Ma=H$, obtained from data in Figs.~\ref{fig:2_paper1}\textbf{c},\textbf{d}. The optimum threshold maximizing the contrast between the two prepared states is the same in both cases. Deviations of the outcome from the intended prepared state are: $8.9\%$ ($1.3\%$) for the ground state, $6.2\%$ ($2.1\%$) for the excited state without (with) conditioning. Therefore, initialization by measurement and postselection increases the readout contrast from $84.9\%$ to $96.6\%$. \textbf{b,}  Readout contrast (purple) and postselected fraction (black) as a function of $\Vthr$. 
\textbf{c,}  Schematics of the readout error model, including the qubit populations in the steady state and at $\tau = 2.4~\us$ after $\Ma$. Only the arrows corresponding to readout errors are shown.
\textbf{d,}  Rabi oscillations of $\QA$ without (empty) and with (full dots) initialization by measurement and postselection. In each case, data are taken by first digitizing $10\,000$ single shots of $\Mb$ into $H$ or $L$, then averaging the results. Error bars on the average values are estimated from a subset of 175 measurements per point. For each angle, 7 randomly-chosen single-shot outcomes are also plotted (black dots at 0 or 1). The visibility of the averaged signal increases upon conditioning $\Mb$ on $\Ma=H$. Figure adapted from Ref.~\citenum{Riste12}.}
\end{figure}

The ability to better discern the qubit states with the JPA-backed readout is quantified by collecting statistics of single-shot measurements. The sequence used to benchmark the readout includes two measurement pulses, $\Ma$ and $\Mb$, each $700~\ns$ long, with a central integration window of $300~\ns$ (Fig.~\ref{fig:2_paper1}\textbf{a}).
Immediately before $\Mb$, a $\pi$ pulse is applied to $\QA$ in half of the cases, inverting the population of ground and excited state (Fig.~\ref{fig:2_paper1}\textbf{b}). We observe a dominant peak for each prepared state, accompanied by a smaller one overlapping with the main peak of the other case.
We hypothesize that the main peak centered at positive voltage corresponds to state $\ket{00}$, and that the smaller peaks are due to residual qubit excitations, mixing the two distributions. To test this hypothesis, we first \hl{digitize} the result of $\Ma$ with a threshold voltage $V_\mathrm{th}$, chosen to maximize the contrast between the cumulative histograms for the two prepared states (Fig.~\ref{fig:2_paper1}\textbf{c}), and assign the value $H (L)$ to the shots falling above (below) the threshold. Then we only keep the results of $\Mb$ corresponding to $\Ma=H$. Indeed, we observe that \hl{postselecting} $91\%$ of the shots  reduces the overlaps from $\sim6$ to $2\%$ and from $\sim9$ to $1\%$ in the $H$ and $L$ regions, respectively (Fig.~\ref{fig:2_paper1}\textbf{d}). This supports the hypothesis of partial qubit excitation in the steady state, lifted by restricting to a subset of measurements where $\Ma$ declares the register to be in $\ket{00}$. Further evidence is obtained by observing that moving the threshold substantially decreases the fraction of postselected measurements without significantly improving the contrast [$\sim+0.1~ (0.2)\%$ keeping $85 ~(13)\%$ of the shots] (Fig.~\ref{fig:3_paper1}\textbf{b}).
Postselection is effective at suppressing the \hl{residual excitation} of any one qubit, since the $\ket{01}$ and $\ket{10}$ distributions are both highly separated from $\ket{00}$, and the probability that both qubits are excited is only $\sim0.2\%$~.

The performance of JPA-backed readout and the effect of initialization by measurement are quantified by the optimum readout contrast, defined as the maximum difference between the cumulative probabilities for the two prepared states (Fig.~\ref{fig:3_paper1}\textbf{a}). Without \hl{initialization}, the use of the JPA gives an optimum contrast of $84.9\%$, a significant improvement over the $26\%$ obtained without the pump tone.
Comparing the deviations from unity contrast without and with initialization, we can extract the parameters for the error model shown in Fig.~\ref{fig:3_paper1}\textbf{c}. The model~\cite{Riste12} takes into account the residual steady-state excitation of both qubits, found to be $\sim4.7\%$ each, and the error probabilities for the qubits prepared in the four basis states. Although the projection into $\ket{00}$ occurs with $99.8\pm0.1\%$ fidelity, this probability is reduced to $98.8\%$ in the time $\tau=2.4~\us$ between $\Ma$ and $\Mb$, chosen to fully deplete the cavity of photons before the $\pi$ pulse preceding $\Mb$. We note that $\tau$ could be reduced by increasing $\kappa$ by at least a factor of two without compromising $T_{1\mathrm{A}}$  by the Purcell effect~\cite{Houck08}. By correcting for partial equlibration during $\tau$, we calculate an actual readout fidelity of $98.1\pm0.3\%$. The remaining infidelity is mainly attributed to qubit relaxation during the integration window.

As a test for readout fidelity, we performed single-shot measurements of a Rabi oscillation sequence applied to $\QA$, with variable amplitude of a resonant $32~\ns$ Gaussian pulse preceding $\Mb$, and using ground-state initialization as described above (Fig.~\ref{fig:3_paper1}\textbf{d}).
The density of discrete dots reflects the probability of measuring $H$ or $L$ depending on the prepared state. By averaging over $\sim10\,000$ shots, we recover the sinusoidal Rabi oscillations without (white) and with (black) ground-state initialization. As expected, the peak-to-peak amplitudes ($85.2$ and $96.7\%$, respectively) equal the optimum readout contrasts in Fig.~\ref{fig:3_paper1}\textbf{a}, within statistical error.

\subsection{Repeated quantum nondemolition measurements}

\begin{figure}
\center
\includegraphics[width=0.7\columnwidth]{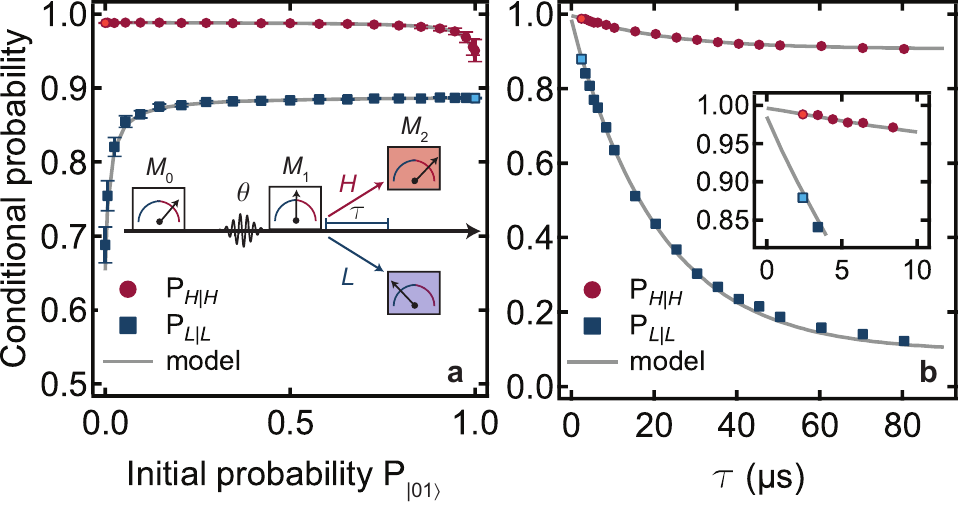}
\caption{\label{fig:4_paper1} \textbf{Projectiveness of the measurement.} {\textbf a,} Conditional probabilities for two consecutive measurements $\Mb$ and $\Mc$, separated by $\tau=2.4~\us$. Following an initial measurement pulse $\Ma$ used for initialization into $\ket{00}$ by the method described, a Rabi pulse with variable amplitude rotates $\QA$ by an angle $\theta$ along the $x$-axis of the Bloch sphere, preparing a state with $P_{\ket{01}} = \sin^2(\theta/2)$. Red (blue): probability to measure $\Mc=H (L)$ conditioned on having obtained the same result in $\Mb$, as a function of the initial excitation  of $\QA$. Error bars are the standard error obtained from 40 repetitions of the experiment, each one having a minimum of 250 postselected shots per point. Deviations from an ideal projective measurement are due to the finite readout fidelity, and to partial recovery after $\Mb$~. The latter effect is shown in \textbf{b}, where the conditional probabilities converge to the unconditioned values,  $ P_{H}=0.91$ and $ P_{L}=0.09$ for $\tau\gg T_1$, in agreement with Fig.~\ref{fig:2_paper1}, taking into account relaxation between the $\pi$ pulse and $\Mc$. Error bars are smaller than the dot size. Figure taken from Ref.~\citenum{Riste12}.}
\end{figure}

In an ideal projective measurement, there is a one-to-one relation between the outcome and the post-measurement state.
We perform repeated measurements to assess the \hl{nondemolition} nature of the readout, following Refs.~\citenum{Lupascu07,Boulant07}.
The correlation between two consecutive measurements, $\Mb$ and $\Mc$, is found to be independent of the initial state over a large range of Rabi rotation  angles $\theta$ (see Fig.~\ref{fig:4_paper1}\textbf{a}).  A decrease in the probabilities occurs when the chance to obtain a certain outcome on $\Mb$ is low (for instance to measure $\Mb=H$ for a state close to $\ket{01}$) and comparable to readout errors or to the partial recovery arising between $\Mb$ and $\Mc$. We extend the readout model of Fig.~\ref{fig:3_paper1}\textbf{c} to include the correlations between each outcome on $\Mb$ and the post-measurement state. The deviation of the asymptotic levels from unity, $P_{H|H}=0.99$ and $P_{L|L}=0.89$, is largely due to recovery during $\tau$, as demonstrated in Fig.~\ref{fig:4_paper1}\textbf{b}. From the model, we extrapolate the correlations for two adjacent measurements, $P_{H|H}(\tau=0)=0.996\pm0.001$ and  $P_{L|L}(\tau=0)=0.985\pm0.002$, corresponding to the probabilities that pre- and post-measurement state coincide. In the latter case, mismatches between the two outcomes are mainly due to qubit relaxation during $\Mc$.
Multiple  measurement pulses, as well as a long pulse, do not have a significant effect on the qubit state, supporting the nondemolition character of  the readout at the chosen power.

Josephson parametric amplification has become a standard technique for the high-fidelity readout of qubits in cQED. Since this experiment and the parallel work in Ref.~\citenum{Johnson12}, projective readout of transmon~\cite{Steffen13, Chow14, Jeffrey14} and flux~\cite{Lin13} qubits has been performed using different varieties of Josephson junction-based amplifiers. 
The technology for these amplifiers continuously evolves to meet the needs of quantum circuits of growing complexity. One approach to high-fidelity readout of multiple qubits is to increase the amplifier bandwidth to include several resonators, each coupled to a distinct qubit~\cite{Groen13}. 
Recent implementations in this direction included Josephson junctions in a transmission line~\cite{OBrien14}, in low-Q resonators~\cite{Mutus14, Eichler14}, or in a circuit realizing a superconducting low-inductance undulatory galvanometer (SLUG)~\cite{Hover14}. Another approach for multi-qubit readout uses dedicated, on-chip Josephson bifurcation amplifiers~\cite{Schmitt14}. 

\section{Digital feedback controllers}
\label{sec:fbcontrollers}
The input to a feedback loop in cQED is the homodyne signal obtained by amplification and demodulation of the qubit-dependent cavity transmission or reflection, as shown above. The response of the \hl{feedback controller} is one or more qubit microwave pulses, which are generated and sent to the device (Fig.~\ref{fig:fbcqed}). This loop has a significant spatial extension, as the qubits sit in the coldest stage of a dilution refrigerator, while the feedback controller is at room temperature. A round trip involves $5-10~\m$ of cable, which translates  to a propagation time of $25-50~\ns$ without accounting for delays due to filters and other microwave components.  
This physical limitation, which would require fast cryogenic electronics to be overcome, is only a small fraction of the total latency. A major source of delay is the \hl{processing time} in the controller, combined with the generation or triggering of the microwave pulses for the conditional qubit rotations. The details of this process depend on the type of controller. We describe the first implementations below.   

\begin{figure*}
\centering
\includegraphics[width=\columnwidth]{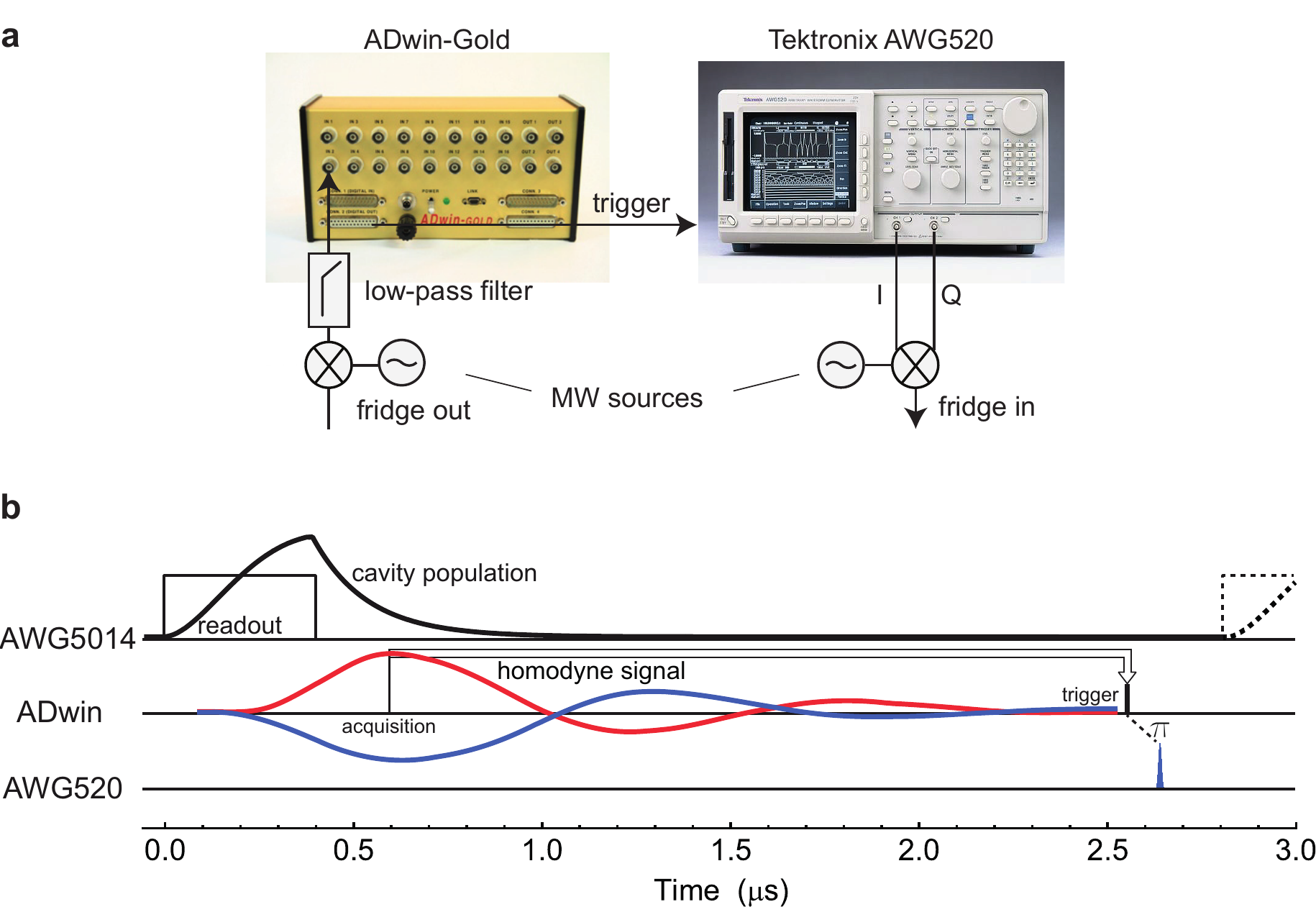}

\caption{\label{fig:s2_paper2} \textbf{Digital feedback loop with an ADwin controller.} \textbf{a,} Schematic of the feedback loop, consisting of an ADwin, sampling the signal, and a Textronix AWG520, conditionally generating a qubit $\pi$ pulse. \textbf{b}, Timings of the feedback loop. The measurement pulse, here $400~\ns$ long, reaches the cavity at $t=0$. 
The ADwin, triggered by an AWG5014, measures one channel of the output homodyne signal (red: qubit in $\ket{0}$, blue: $\ket{1}$), delayed by $\sim200~\ns$ due to a low-pass filter at its input side. After comparison of the measured voltage at $t=0.6~\us$ to the reference threshold, the AWG520 is conditionally triggered at $t=2.54~\us$, resulting in a $\pi$ pulse 
reaching the cavity at $2.62~\us$. 
Figure adapted from Ref.~\citenum{Riste12b}.}
\end{figure*}

The first realization of a digital feedback controller used commercial components for data sampling, processing, and conditional operations~\cite{Riste12b}. The core of the controller is an \hl{ADwin-Gold}, a processor with
a set of analog inputs and configurable analog and digital outputs. The ADwin samples the readout signal once, at a set delay following a trigger from an arbitrary waveform generator (Tektronix AWG5014). This delay is optimized to maximize readout fidelity. A routine determines the optimum threshold for digitizing the readout signal. This voltage is then used to assign $H$ or $L$ to the measurement. For the reset function in section~\ref{sec:reset}, the ADwin triggers another arbitrary waveform generator (Tektronix AWG520) to produce a $\pi$ pulse when the outcome is $L$. Pulse timings and signal delays in the feedback cycle are illustrated in Fig.~\ref{fig:s2_paper2}. The total time between start of the measurement and end of the feedback pulse is $\approx 2.6~\us$, mainly limited by the processing time of the ADwin. 

\begin{figure*}
\centering 
\includegraphics[width=\columnwidth]{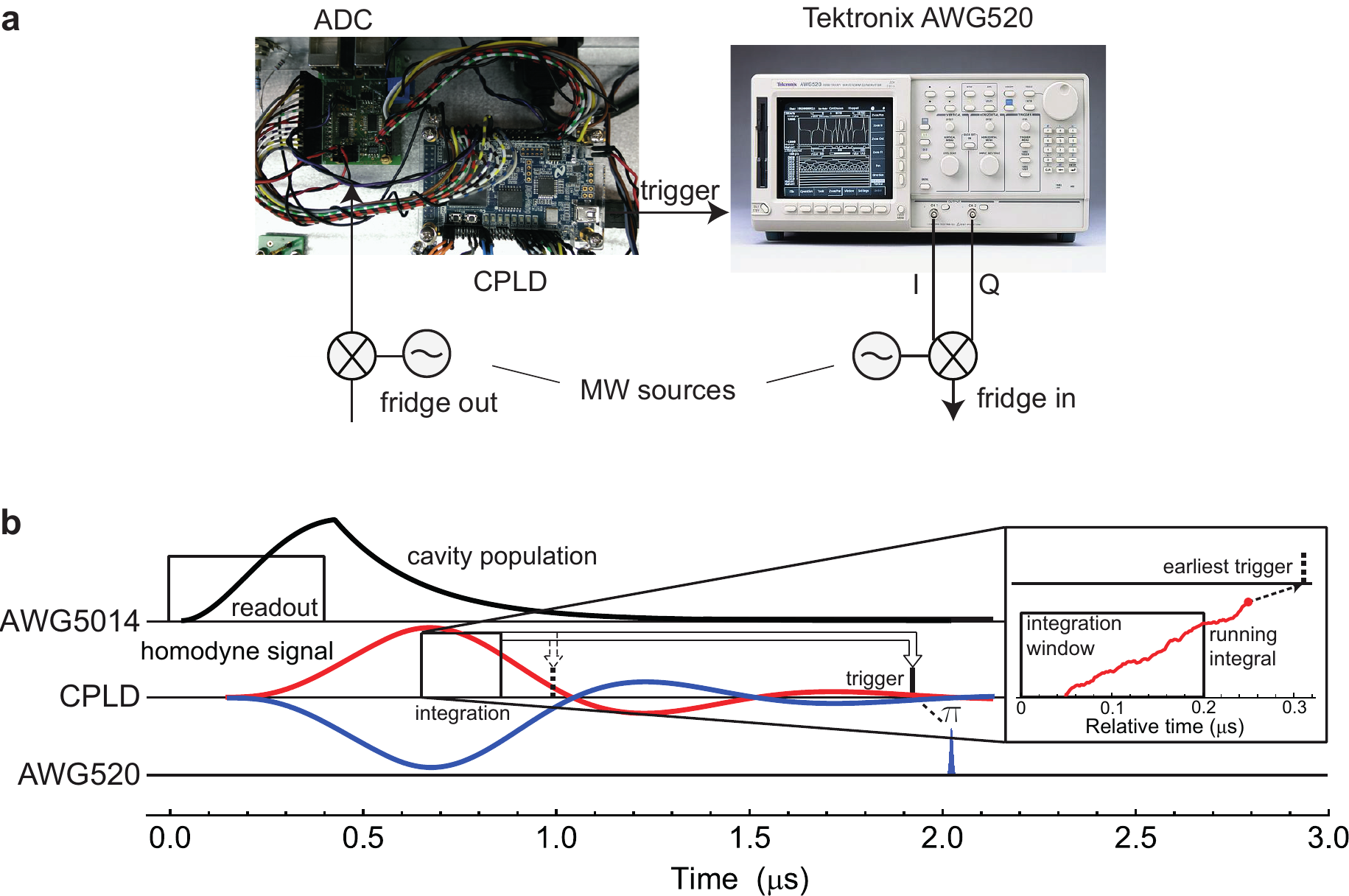}
\caption{\label{fig:2_book} \textbf{Digital feedback loop with a CPLD-based controller.} \textbf{a}, Schematics of the feedback loop, with an ADC and a CPLD (or FPGA) board replacing the ADwin in Fig.~\ref{fig:s2_paper2}. \textbf{b,} Timings of the feedback loop. The CPLD samples the signal at every clock cycle ($10~\ns$) and then integrates it over a window set by a marker of an AWG5014. The internal delay of the CPLD breaks down into the analog-to-digital conversion ($60~\ns$) and the processing to compare the integrated signal to a calibrated threshold, determining the binary output ($50~\ns$). These timings are multiples of the clock (reduced to $4~\ns$ in a recent FPGA-based implementation~\cite{reportGarrido14}). The total delay in Ref.~\cite{Riste13} is increased to $2~\us$ to let the cavity return to the ground state before the conditional $\pi$ pulse.} 

\end{figure*}

To shorten the loop time, our second generation of digital feedback used a complex programmable logic device (\hl{CPLD}, Altera MAX V), acquiring the signal following a $8$-bit ADC, in place of the ADwin. This home-assembled feedback controller offers two advantages over the first: a programmable integration window and a response time of $0.11~\us$ (Fig.~\ref{fig:2_book}), an order of magnitude faster than the ADwin. 
As the feedback response time is now comparable or faster than the typical cavity decay time, active depletion of the cavity~\cite{McClure15} will be required to take full advantage of the CPLD speed and further shorten the feedback loop. 

\begin{figure*}
\centering 
\includegraphics[width=0.8\columnwidth]{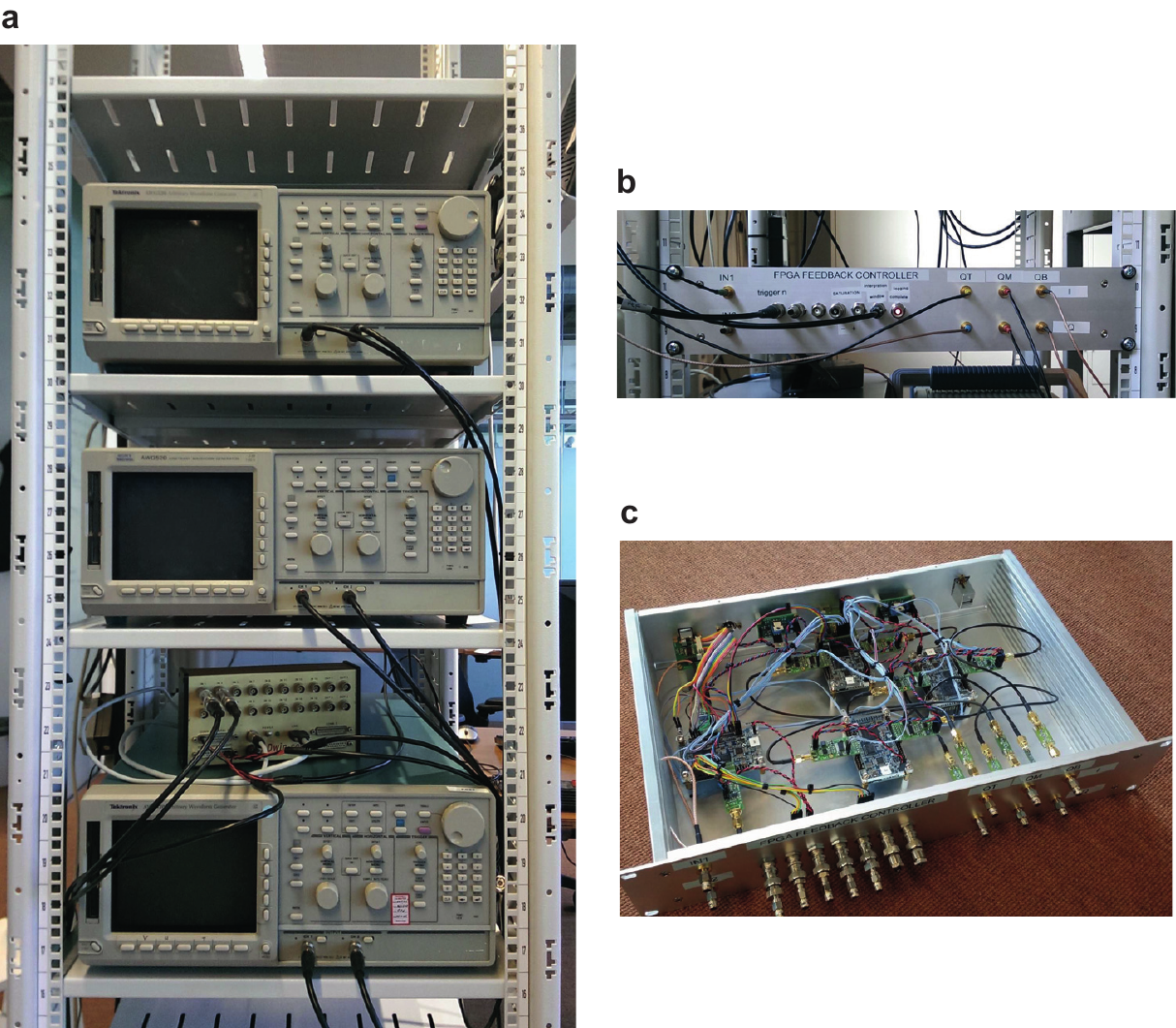}
\caption{\label{fig:fbcomparison} \textbf{Hardware comparison for feedback control in the bit-flip code.} The bit-flip code requires a two-bit digital feedback, acting on three qubits. Scaling the system in Fig.~\ref{fig:s2_paper2} would take an AWG520 for each qubit (\textbf{a}). 
A recent implementation~\cite{reportGarrido14} performs readout signal processing and pulse generation on FPGA boards, resulting in the compact controller shown in \textbf{b}, \textbf{c}.} 
\end{figure*}

Further developments in the feedback controller replaced the CPLD with a field-programmable-gate-array (\hl{FPGA}) to increase the on-board memory and enable more complex signal processing. For example, the FPGA allows different weights for the measurement record and  maximal correlation with the qubit evolution. A FPGA-based controller has also been employed for digital feedback at ETH Zurich~\cite{Steffen13}. Recent developments at TU Delft and at Yale~\cite{Ofek15} include the pulse generation on a FPGA board, eliminating the need of an additional AWG. For comparison, Fig.~\ref{fig:fbcomparison} shows the setup that would be required for the 3-qubit repetition code~\cite{Mermin07} using our first generation of feedback (\textbf{a}) and the most recent one based on FPGAs (\textbf{b}, \textbf{c}).

\section{Fast qubit reset based on digital feedback} 

\begin{figure}
\centering
\includegraphics[width=0.5\columnwidth]{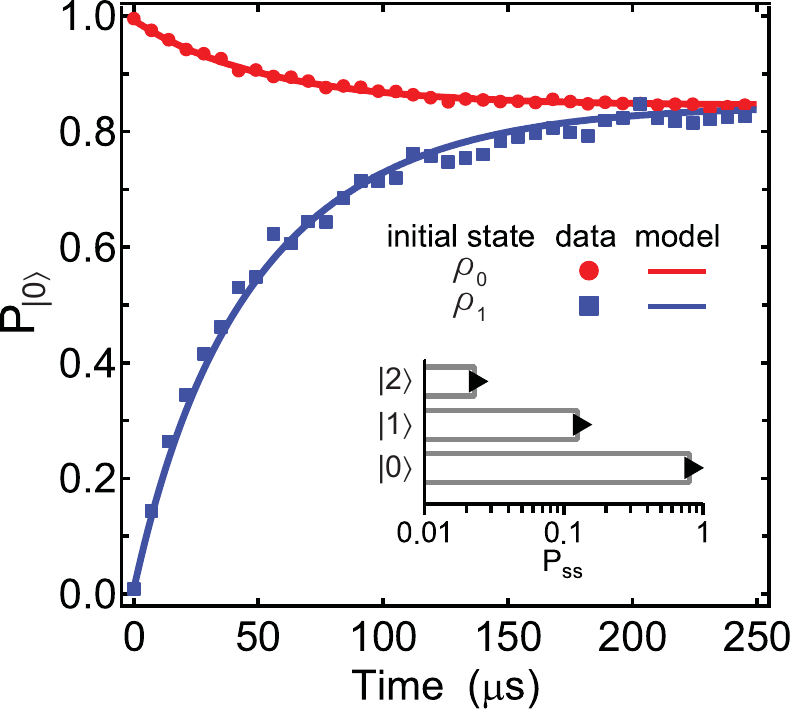}
\caption{\label{fig:1_paper2} \textbf{Transmon equilibration to steady state.} Time evolution of the ground-state population $\Pg$ starting from states $\rho_0$ and $\rho_1$ (notation defined in the text). Solid curves are the best fit to Eq.~\eqref{eq:rates_paper2}, giving the inverse transition rates $\Gzo^{-1 }= 50\pm2~\us, \Got^{-1} = 20\pm2~\us, \Goz^{-1} = 324\pm32~\us, \Gto^{-1} = 111\pm25~\us$. From the steady-state solution, we extract residual excitations $P_{\ket{1},\mathrm{ss}}=13.1\pm0.8\%,P_{\ket{2},\mathrm{ss}}=2.4\pm0.4\%$. Inset: steady-state population distribution (bars). Markers correspond to a Boltzmann distribution with best-fit  temperature $127~\mK$, significantly higher than the dilution refrigerator base temperature ($15~\mK).$ Figure taken from Ref.~\citenum{Riste12b}.}
\end{figure}

\subsection{Passive qubit initialization to steady state}

\begin{figure}
\centering
\includegraphics[width=0.55\columnwidth]{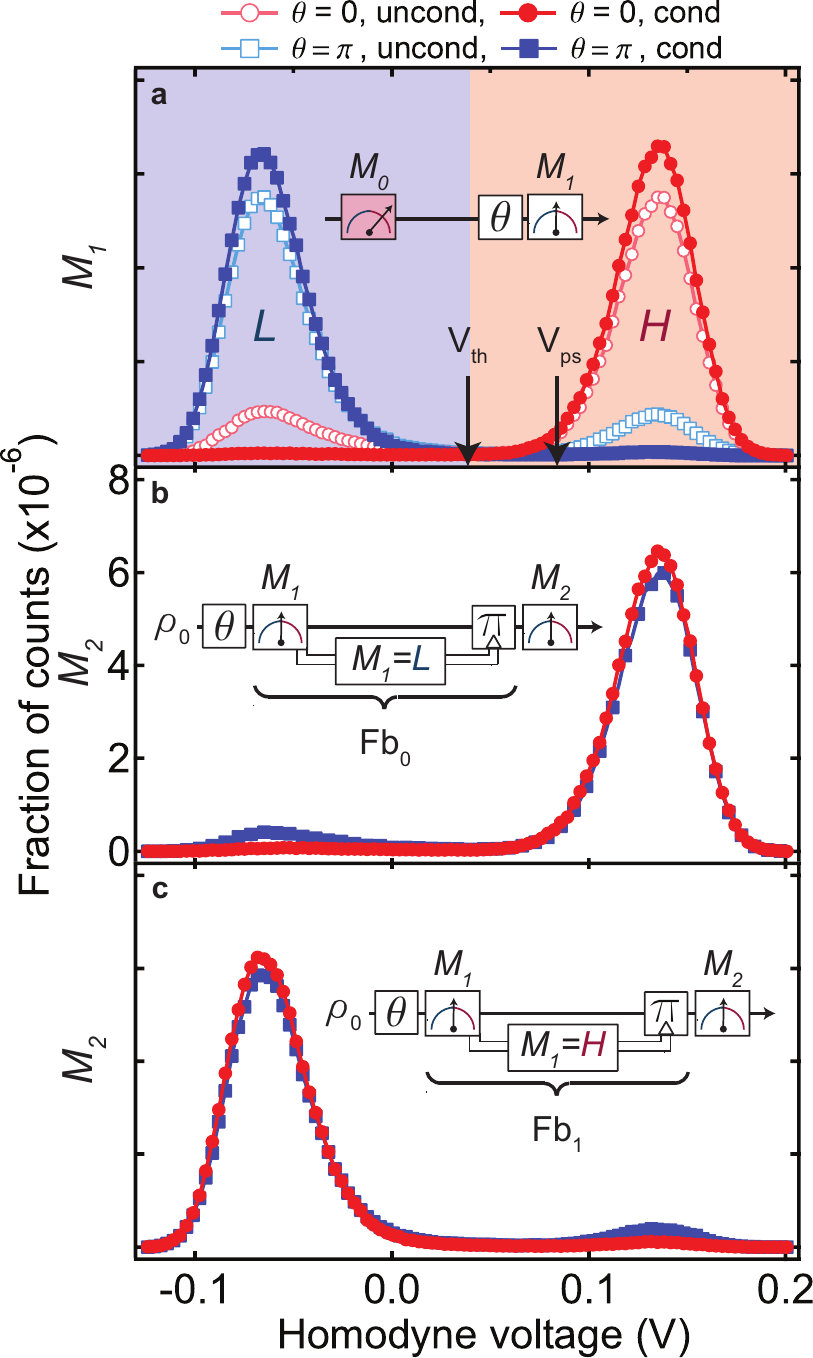}
\caption{\label{fig:2_paper2} \textbf{Reset by measurement and feedback.} \textbf{a,} Before feedback: histograms of $300\,000$ shots of $\Mb$, with (squares) and without (circles) inverting the qubit population with a $\pi$ pulse. Each shot is obtained by averaging the homodyne voltage over the second half ($200~\ns$) of a readout pulse. $H$ and $L$ denote the two possible outcomes of $\Mb$, digitized with the threshold $V_\mathrm{th}$, maximizing the contrast, analogously to section~\ref{sec:readout}.  Full (empty) dots indicate (no) postselection on $\Ma>\Vps$. This protocol~ is used to prepare $\rho_0$ and $\rho_1$, which are the input states for the feedback sequences in \textbf{b} and \textbf{c}.
\textbf{b,} After feedback: histograms of $\Mc$  after applying the feedback protocol $\Fbg$, which triggers a $\pi$ pulse when $\Mb=L$. Using this feedback,  $\sim99\%$ $(92\%)$ of measurements digitize to $H$ for $\theta=0~(\pi)$, respectively. \textbf{c,} Feedback with opposite logic $\Fbe$ preparing the excited state. In this case, $\sim 98\%$ $(94\%)$ of measurements digitize to $L$ for $\theta=0~(\pi)$.  Figure taken from Ref.~\citenum{Riste12b}.}
\end{figure}

Our first application of feedback is qubit initialization, also known as reset~\cite{DiVincenzo00}. The ideal reset for QIP is \hl{deterministic} (as opposed to heralded or postselected, see previous section) and fast compared to qubit coherence times. Obviously, the passive method of waiting several times $T_1$ does not meet the speed requirement. Moreover, it can suffer from residual steady-state qubit excitations~\cite{Corcoles11,Johnson12,Riste12,Vijay12}, whose cause in cQED remains an active research area. The drawbacks of \hl{passive initialization} are evident for our qubit, whose ground-state population $\Pg$ evolves from states  $\rho_0$ and $\rho_1$ as shown in Fig.~\ref{fig:1_paper2}. With $\rho_0$ and $\rho_1$ we indicate our closest realization ($\sim99\%$ fidelity) of the ideal pure states $\ket{0}$ and $\ket{1}$. $\Pg$ at variable time after preparation is obtained by comparing the average readout homodyne voltage to calibrated levels~, as in standard  three-level tomography~\cite{Thew02,Bianchetti09}.
These populations dynamics are captured by a master equation model for a three-level system:
\begin{equation}
\label{eq:rates_paper2}
\
\left(\begin{array}{c}
\dot{P}_\mathrm{\ket{0}}\\
\dot{P}_\mathrm{\ket{1}}\\
\dot{P}_\mathrm{\ket{2}}
\end{array}\right)=
\left( \begin{array}{ccc}
-\Goz & \Gzo & \phantom{-}0 \\
\phantom{-}\Goz & -\Gzo-\Gto & \phantom{-}\Got \\
\phantom{-}0 & \Gto & -\Got
\end{array} \right)
\left(\begin{array}{c}
P_\mathrm{\ket{0}}\\
P_\mathrm{\ket{1}}\\
P_\mathrm{\ket{2}}
\end{array}\right).
\end{equation}

The best fit to the data gives the qubit relaxation time $T_1=1/\Gzo=50\pm2~\us$ and the asymptotic $15.5\%$ residual total excitation.

\subsection{Qubit reset based on digital feedback}
\label{sec:reset}

\begin{figure}
\centering
\includegraphics[width=0.65\columnwidth]{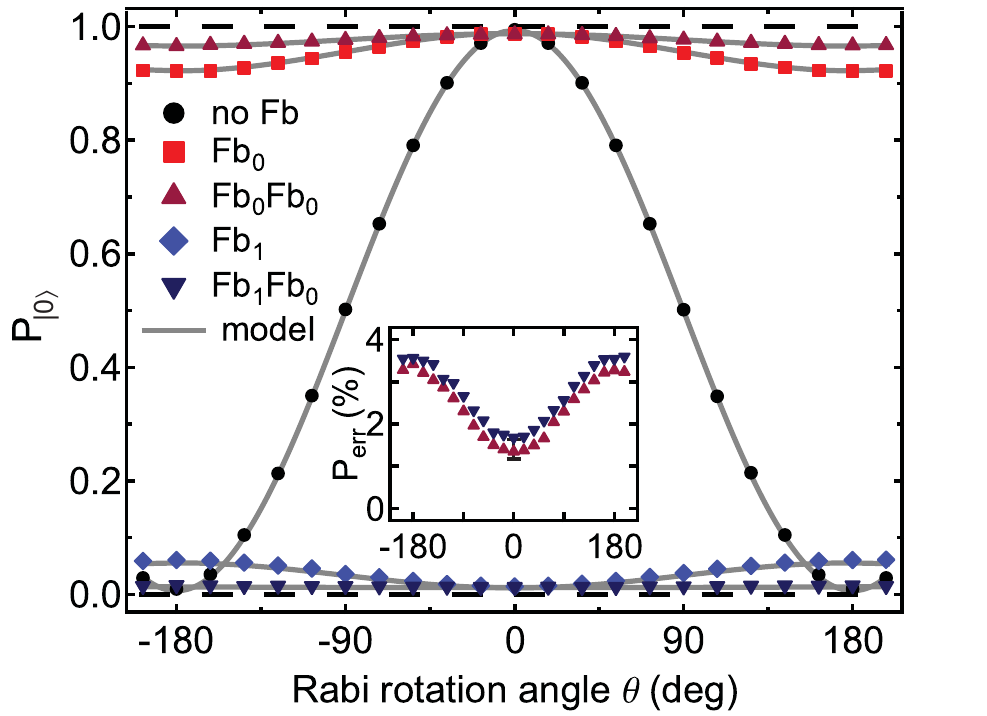}
\caption{\label{fig:3_paper2} \textbf{Deterministic reset from any qubit state.}
Ground-state population $\Pg$ as a function of the initial state $\rho_\theta$, prepared by coherent rotation after initialization in $\rho_0$, as in Fig.~\ref{fig:2_paper2}. The cases shown are: no feedback (circles), $\Fbg$ (squares), $\Fbe$ (diamonds), twice $\Fbg$ (upward triangles), and $\Fbg$ followed by $\Fbe$ (downward triangles). The vertical axis is calibrated with the average measurement outcome for the reference states $\rho_0, \rho_1$, and corrected for imperfect state preparation. The curve with no feedback has a visibility of $99\%$, equal to the average preparation fidelity. Each experiment is averaged over $300\,000$ repetitions. Inset: error probabilities for two rounds of feedback, defined as $1-P_{\ket{t}}$, where $\ket{t}\in\{0,1\}$ is the target state. The systematic $\sim0.3\%$ difference between the two cases is attributed to error in the $\pi$ pulse preceding the measurement of $\Pe$ following $\Fbe$.  Curves: model including readout errors and equilibration (section~\ref{sec:fbbasics}).}
\end{figure}

Previous approaches to accelerate qubit equilibration include coupling to dissipative resonators~\cite{Reed10b} or two-level systems~\cite{Mariantoni11}. However, these are also susceptible to spurious excitation, potentially inhibiting complete qubit relaxation. Feedback-based reset circumvents the equilibration problem by not relying on coupling to a dissipative medium. Rather, it works by projecting the qubit with a measurement ($\Mb$, performed by the controller) and conditionally applying a $\pi$ pulse to drive the qubit  to a targeted basis state (Fig.~\ref{fig:2_paper2}).
A final measurement ($\Mc$) determines the qubit state immediately afterwards. 
In both measurements, the result is digitized into levels $H$ or $L$, associated with $\ket{0}$ and $\ket{1}$, respectively. The digitization threshold voltage $\Vthr$ maximizes the readout fidelity at $99\%$.
The $\pi$ pulse  is conditioned on $\Mb=L$ to target $\ket{0}$ (scheme $\Fbg$) or on $\Mb=H$ to target $\ket{1} (\Fbe)$.
In a QIP context, reset is typically used to reinitialize a qubit following measurement, when it is in a computational basis state. Therefore, to benchmark the reset protocol, we first quantify its action on $\rho_0$ and $\rho_1$.
This step is accomplished with a preliminary  measurement $\Ma$ (initializing the qubit in $\rho_0$ by postselection), followed by a calibrated pulse resonant on the transmon $\zotrans$ transition to prepare $\rho_1$.
The overlap of the $\Mc$ histograms with the targeted region ($H$ for $\Fbg$ and $L$ for $\Fbe$) averages at $96\%$, indicating the success of reset. Imperfections are more evident for $\theta=\pi$ and mainly due to equilibration of the transmon during the feedback loop. A detailed error analysis is presented below. We emphasize that qubit initialization by postselection  
is here only used to prepare nearly pure states useful for characterizing the feedback-based reset, which is deterministic.

\subsection{Characterization of the reset protocol}

An ideal reset function prepares the same pure qubit state regardless of its input. To fully quantify the performance of our reset scheme, we measure its effect on our closest approximation to superposition states $\ket{\theta}=\cos(\theta/2)\ket{0}+\sin({\theta/2})\ket{1}$.
Without feedback, $\Pg$ is trivially a sinusoidal function of $\theta$, with near unit contrast. Feedback highly suppresses the Rabi oscillation, with $\Pg$ approaching the ideal value 1 (0) for $\Fbg$  $(\Fbe)$ for any input state. However, a dependence on $\theta$ remains, with $\Perr=1-\Pg$ for $\Fbg$ ($1-\Pe$ for $\Fbe$) ranging from $1.2\%$ $(1.4\%)$ for $\theta=0$ to $7.8\%$ ($8.4\%$) for $\theta = \pi$. The remaining errors are discussed in section~\ref{sec:fbbasics}. 
From Eqs.~\eqref{eq:fb}, using the best-fit $\Gamma_{ij}$ and $\tlat = 2.4~\us$, errors due to equilibration sum to $0.7\%$ $(6.9\%)$ for $\theta=0$ $(\pi)$, while readout errors account for the remaining $0.4\%$ $(1.4\%)$. 
In agreement with these values, concatenating two feedback cycles suppresses the error for $\theta  = \pi$ to $3.4\%$, while there is no benefit for $\theta = 0$ ($1.3\%$). 

\begin{figure}
\centering
\includegraphics[width=0.75\columnwidth]{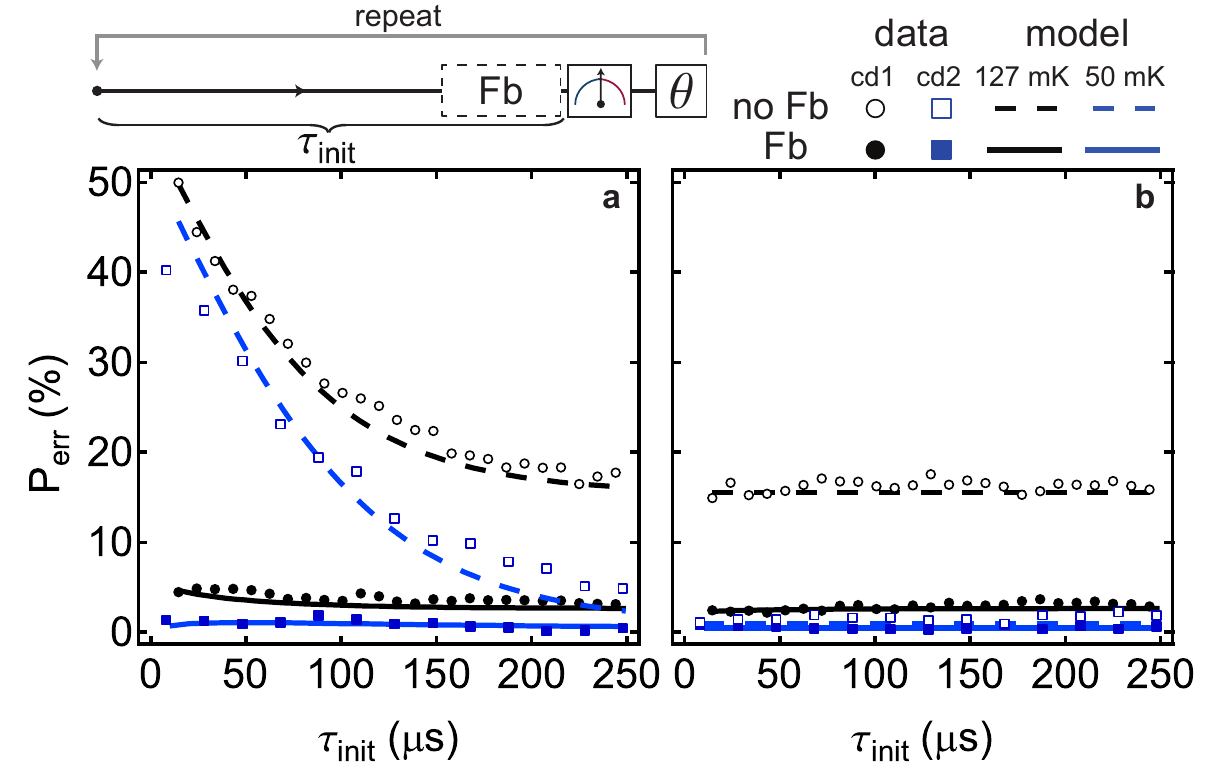}
\caption{\label{fig:4_paper2} \textbf{Fast qubit reset.} Initialization errors as a function of initialization time $\tinit$ under looped execution of a simple experiment leaving the qubit ideally in $\ket{1}$ (\textbf{a,} measurement and $\pi$ pulse) or $\ket{0}$ (\textbf{b,} measurement only). Empty symbols: initialization by waiting (no feedback). Solid symbols: initialization by feedback, with three rounds of $\Fbg$ and a $\pi$ pulse on the $1\leftrightarrow2$ transition. Two data sets correspond to two different cooldowns: the one corresponding to Figs.~\ref{fig:1_paper2}-\ref{fig:3_paper2} (black) and a following one with improved thermalization (blue).  
Curves correspond to a master equation simulation assuming perfect pulses and measured transition rates $\Gamma_{ij}$ (dashed, no feedback; solid, triple $\Fbg$ with a $\pi$ pulse on $1\leftrightarrow2$). Feedback reset successfully bounds the otherwise exponential accruement of $\Perr$ in case \textbf{a} as $\tinit\rightarrow0$. The reduction of $\Perr$ in \textbf{b} reflects the cooling of the transmon by feedback (see text for details). Figure adapted from Ref.~\citenum{Riste12b}.}
\end{figure}

\subsection{Speed-up enabled by fast reset}

The key advantage of reset by feedback is the ability to ready a qubit for further computation \hl{fast} compared to coherence times available in 3D cQED~\cite{Paik11,Rigetti12}. This will be important, for example, when refreshing ancilla qubits in multi-round error correction~\cite{Schindler11}. We now show that reset suppresses the accumulation of initialization error when a simple experiment is repeated with decreasing in-between time $\tinit$.  The simple sequence in Fig.~\ref{fig:4_paper2} emulates an algorithm that leaves the qubit in $\ket{1}$ [case (a)] or $\ket{0}$ [case (b)]. A measurement pulse follows $\tinit$ to quantify the initialization error $\Perr$. Without feedback, $\Perr$ in case (a) grows exponentially as $\tinit\to0$. This accruement of error, due to the rapid succession of $\pi$ pulses, would occur even at zero temperature, where residual excitation would vanish (i.e., $\Gamma_{i+1,i}=0$), in which case $\Perr \to 50\%$ as $\tinit\rightarrow0$. In case (b), $\Perr$ matches the total steady-state excitation for all $\tinit$. Using feedback significantly improves initialization for both long and short $\tinit$. For $\tinit \gg T_1$, feedback suppresses $\Perr$ from the $16\%$ residual excitation to $3\%$ (black symbols and curves)\footnote{We note that $\Pe\approx\Pee=1.6\%$ is a non-thermal distribution.}, cooling the transmon. Crucially, unlike passive initialization, reset by feedback is also effective at short $\tinit$, where it limits the otherwise exponential accruement of error in (a), bounding $\Perr$ to an average of $3.5\%$ over the two cases. 
Our scheme combines three rounds of $\Fbg$ with a pulse on the $1\leftrightarrow2$  transition before the final $\Fbg$ to partially counter leakage to the second excited state, which is the dominant error source [see Eq.~\eqref{eq:fb}]. The remaining leakage is proportional to the average $\Pe$, which slightly increases in \textbf{a} and decreases in \textbf{b} as $\tinit\to0$.  
In a following cooldown, with improved thermalization and a faster feedback loop (Fig.~\ref{fig:2_book}), reset  
constrained $\Perr\lesssim1\%$ (blue), quoted as the fault-tolerance threshold for initialization in modern error correction schemes~\cite{Wang11}. In addition to the near simultaneous implementation at ENS~\cite{CampagneIbarcq13}, similar implementations of qubit reset have followed at Yale~\cite{Ofek15} and at Raytheon BBN Technologies using a FPGA-based feedback controller. 

\section{Deterministic entanglement by parity measurement and feedback}
\label{sec:ebm}
In this section, we extend the use of digital feedback to a multi-qubit experiment, targeting the deterministic generation of \hl{entanglement by measurement}. We first turn the cavity into a \hl{parity meter} to measure the joint state of two coupled qubits. By carefully engineering the cavity-qubit dispersive shifts, we make the cavity transmission only sensitive to the excitation parity, but unable to distinguish states within each parity. Binning the final states on the parity result generates an entangled state in either case, with up to $88\%$ fidelity to the closest Bell state. Integrating the demonstrated feedback control in the parity measurement, we turn the entanglement generation from \hl{probabilistic} to \hl{deterministic}.  
\subsection{Two-qubit parity measurement}
In a two-qubit system, the ideal  parity measurement transforms an unentangled superposition state $\kpsii = (\ket{00}+\ket{01}+\ket{10}+\ket{11})/2$ into  Bell states
\begin{eqnarray}
\label{eq:paper5_Bell}
&&\OddBellp = \frac{1}{\sqrt{2}}(\ket{01}+\ket{10})\,\,\, \mathrm{and} \,\,\, \EvenBellp = \frac{1}{\sqrt{2}}(\ket{00}+\ket{11})
\end{eqnarray}
for odd and even outcome, respectively. 
Beyond generating entanglement between non-interacting qubits~\cite{Ruskov03, Trauzettel06, Ionicioiu07, Williams08b, Haack10}, parity measurements allow deterministic two-qubit gates~\cite{Beenakker04, Engel05} and play a key role as syndrome detectors in quantum error correction~\cite{Nielsen00,Ahn02}. 
A heralded parity measurement has been recently realized for nuclear spins in diamond~\cite{Pfaff12}. 
By minimizing measurement-induced decoherence at the expense of single-shot fidelity, highly entangled states were generated with $3\%$ success probability. 
Here, we realize the first solid-state parity meter that produces entanglement with unity probability.

\begin{figure*}
\centering
\includegraphics[width=\columnwidth]{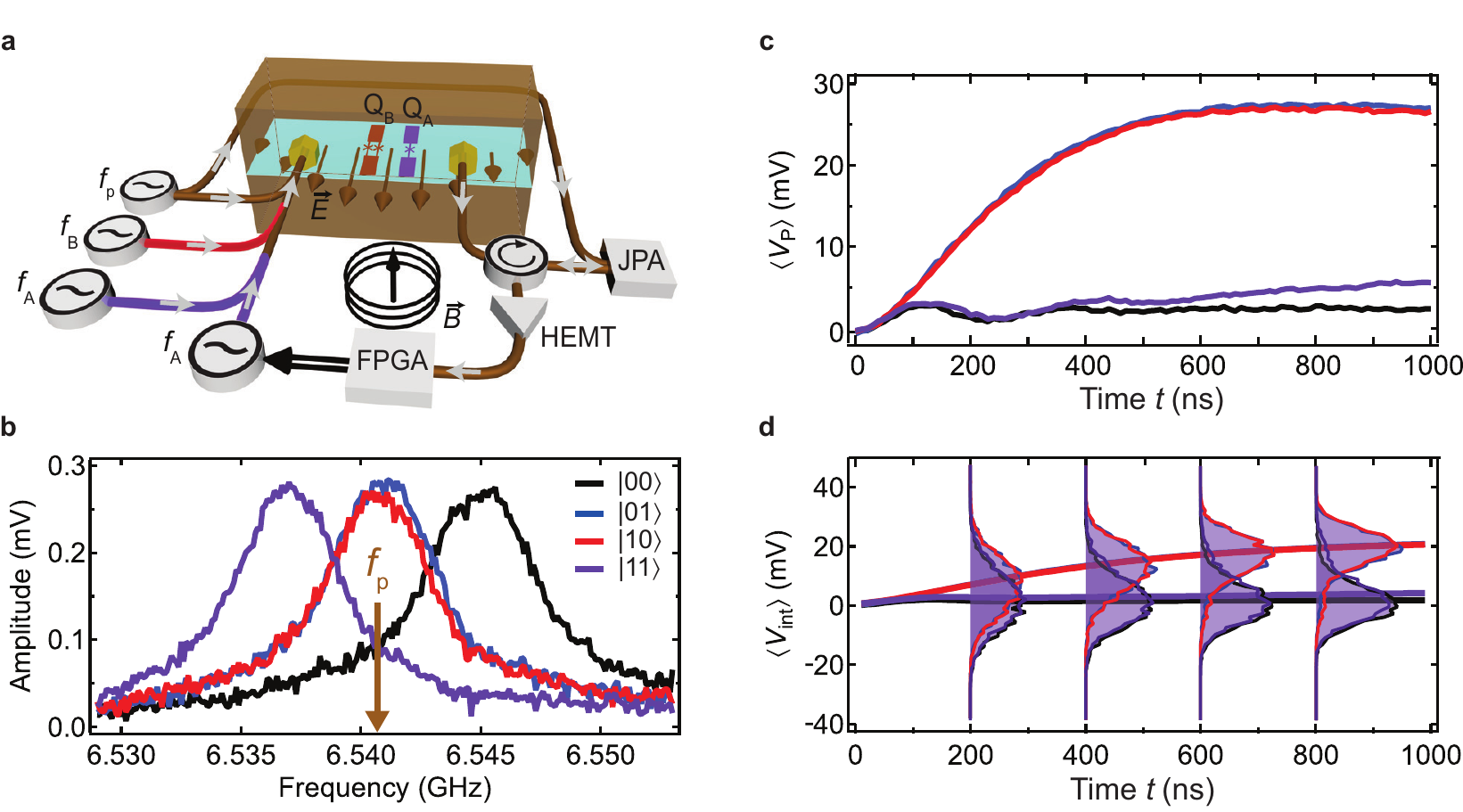}
\caption{\label{fig:1_paper5} \textbf{Cavity-based two-qubit parity readout in cQED.}  \textbf{a,} Simplified diagram of the experimental setup. Single- and double-junction transmon qubits ($\QA$ and $\QB$, respectively) dispersively couple to the fundamental mode of a 3D copper cavity enclosing them. Parity measurement is performed by homodyne detection of the qubit state-dependent cavity response~\cite{Blais04} using a JPA~\cite{Castellanos-Beltran08}. Following further amplification at $4~\K$ (HEMT) and room temperature, the signal is demodulated and integrated. A FPGA controller closes the feedback  loop that achieves deterministic entanglement by parity measurement (Fig.~\ref{fig:4_paper5}).  \textbf{b,} Matching of the dispersive cavity shifts realizing a parity measurement.  
 \textbf{c,} Ensemble-averaged homodyne response $\avg{\Vp}$ for qubits prepared in the four computational basis states. 
 \textbf{d,} Curves: corresponding ensemble averages of the running integral $\avg{\Vint}$ of $\avg{\Vp}$ between $\ti=0$ and $\tf=t$. Single-shot histograms ($5\,000$ counts each) of $\Vint$ are shown in $200~\ns$ increments. Figure adapted from Ref.~\citenum{Riste12b}. 
} 
\end{figure*}

\subsection{Engineering the cavity as a parity meter}

Our parity meter realization exploits the dispersive regime~\cite{Blais04} in two-qubit cQED. Qubit-state dependent shifts of a cavity resonance (here, the fundamental of a 3D cavity enclosing transmon qubits $\QA$ and $\QB$) allow joint qubit readout by homodyne detection of an applied microwave pulse transmitted through the cavity (Fig.~\ref{fig:1_paper5}\textbf{a}).   
The temporal average $\Vint$ of the homodyne response $\Vp(t)$ over the time interval $[\ti,\tf]$ 
constitutes the measurement needle, with expectation value
\[
\langle \Vint \rangle = \tr(\Ojoint \rho),
\]
where $\rho$ is the two-qubit density matrix and the observable $\Ojoint$ has the general form
\[
\Ojoint = \beta_0  + \betaA   \szA + \betaB \szB  + \betaBA \szB  \szA.
\]
The coefficients $\beta_0$, $\betaA$, $\betaB$, and $\betaBA$  depend on the strength $\epsp$, frequency $\fp$ and duration $\tp$ of the measurement pulse, the cavity linewidth $\kappa$, and the frequency shifts $2\ChiA$ and $2\ChiB$ of the fundamental mode when $\QA$ and $\QB$ are individually excited from $\ket{0}$ to $\ket{1}$. 
The necessary condition for realizing a parity meter is $\betaA=\betaB=0$ ($\beta_0$ constitutes a trivial offset). A simple approach~\cite{Hutchison09,Lalumiere10}, pursued here, is to set $\fp$ to the average of the resonance frequencies for the four computational basis states  $\ket{ij}$ ($i,j\in\{0,1\}$) and to match $\ChiA=\ChiB$. 
We engineer this matching by targeting specific qubit transition frequencies $\fa$ and $\fb$ below and above the fundamental mode during fabrication and using an external magnetic field to fine-tune $\fb$ in situ. We align $\ChiA$ to $\ChiB$ to within $\sim0.06\,\kappa=2\pi \times 90~\kHz$ (Fig.~\ref{fig:1_paper5}\textbf{b}). 
The ensemble-average $\avg{\Vp}$ confirms nearly identical high response for odd-parity computational states $\ket{01}$ and $\ket{10}$, and nearly identical low response for the even-parity $\ket{00}$ and $\ket{11}$ (Fig.~\ref{fig:1_paper5}\textbf{c}). The transients observed are consistent with the independently measured $\kappa$, $\ChiA$ and $\ChiB$ values, and the $4~\MHz$ bandwidth of the JPA at the front end of the output amplification chain. Single-shot histograms (Fig.~\ref{fig:1_paper5}\textbf{d}) demonstrate the increasing ability of $\Vint$ to discern states of different parity as $\tf$ grows (keeping $\ti=0$), and its inability to discriminate between states of the same parity.  The histogram separations at $\tf=400~\ns$ give $|\betaA|,|\betaB| < 0.02~|\betaBA|$. 

\begin{figure}
\centering
\includegraphics[width=0.55\columnwidth]{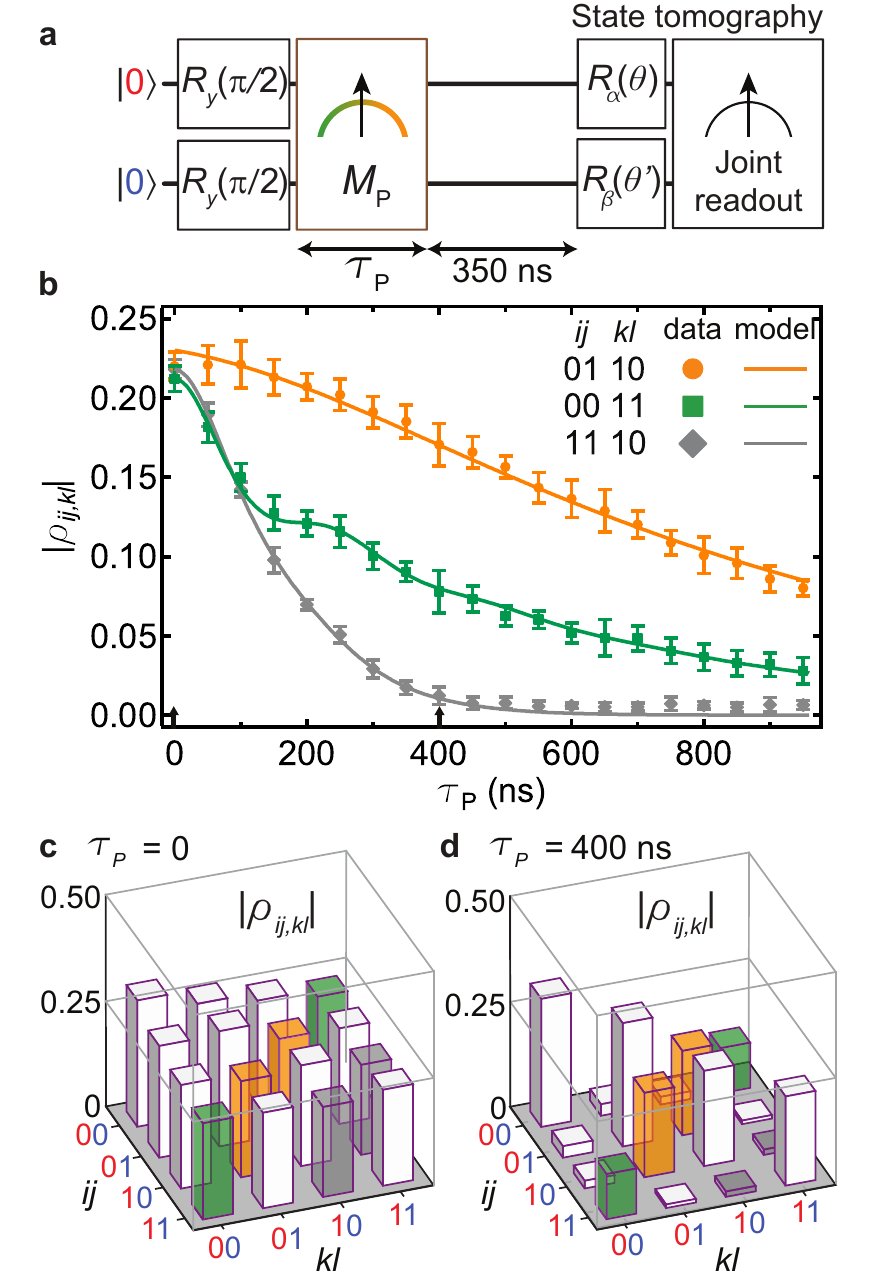}
\caption{\label{fig:2_paper5} \textbf{Unconditioned two-qubit evolution under continuous parity measurement.}  \textbf{a,} Pulse sequence including preparation of the qubits in the maximal superposition state $\rhoi=\kpsii\bpsii$, parity measurement and tomography of the final two-qubit state $\rho$ using joint readout.  \textbf{b,} Absolute coherences $|\rmixed|$, $|\rodd|$, $|\reven|$ following a parity measurement with variable duration $\tp$. Free parameters of the model are the steady-state photon number on resonance $\nss = 2.5\pm 0.1$, the difference $(\ChiA-\ChiB)/\pi=235\pm4~\kHz$, and the absolute coherence values at $\tp=0$ to account for few-percent pulse errors in state preparation and tomography pre-rotations. Note that the frequency mismatch differs from that in Fig.~\ref{fig:1_paper5}\textbf{b} due to its sensitivity to measurement power.  \textbf{c, d,} Extracted density matrices for $\tp=0$ (\textbf{c}) and $\tp=400~\ns$ (\textbf{d}), by which time coherence across the parity subspaces (grey) is almost fully suppressed, while coherence persists within the odd-parity (orange) and even-parity (green) subspaces. Error bars correspond to the standard deviation of $15$ repetitions. Figure taken from Ref.~\citenum{Riste13b}.}
\end{figure}

\subsection{Two-qubit evolution during parity measurement}
Moving beyond the description of the measurement needle, we now investigate the \hl{collapse} of the two-qubit state during parity measurement. We prepare the qubits in the maximal superposition state $\kpsii=\frac{1}{2}\left(\ket{00}+\ket{01}+\ket{10}+\ket{11}\right)$, apply a parity measurement pulse for $\tp$, and perform tomography of the final two-qubit density matrix $\rho$ with and without conditioning on $\Vint$ (Fig.~\ref{fig:2_paper5}\textbf{a}). We choose a weak parity measurement pulse exciting $\nss=2.5$ intra-cavity photons on average in the steady-state, at resonance. A delay of $3.5/\kappa=350~\ns$ is inserted to deplete the cavity of photons before performing  tomography. The tomographic joint readout is also carried out at $\fp$, but with $14~\dB$ higher power, at which the cavity response is weakly nonlinear and sensitive to both single-qubit terms and two-qubit correlations ($\betaA\sim\betaB\sim\betaBA$, as required for tomographic reconstruction~\cite{Filipp09}.

\begin{figure*}
\centering
\includegraphics[width=\columnwidth]{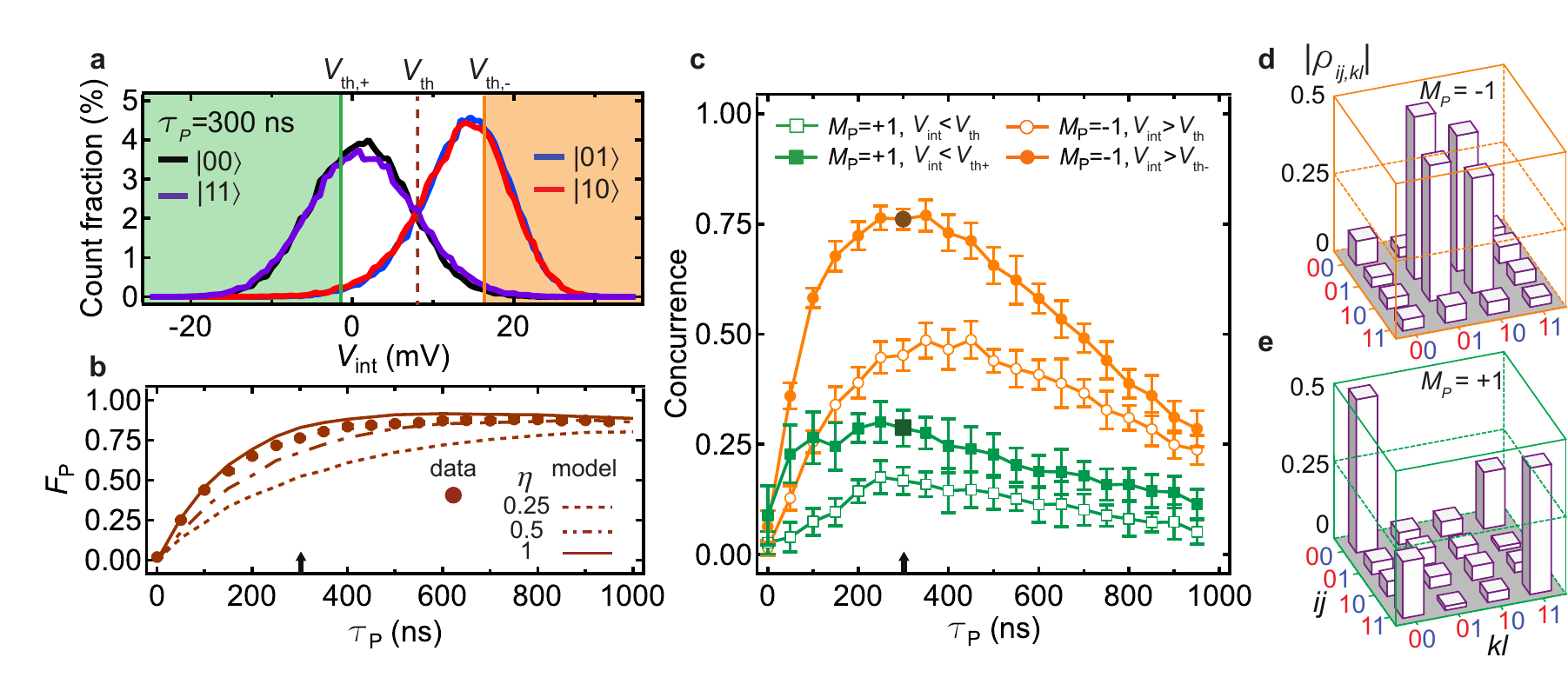}
\caption{\label{fig:3_paper5} \textbf{Probabilistic entanglement generation by postselected parity measurement.}  \textbf{a,} Histograms of $\Vint$ ($\tp=300~\ns$) for the four computational states.  The results are digitized into $\Mp=1 (-1)$ for $\Vp$ below (above) a chosen threshold.   \textbf{b,} Parity readout fidelity $\Fp$ as a function of $\tp$. We define $\Fp=1-\epse-\epso$, with $\epse=p(\Mp=-1|\mathrm{even})$ the readout error probability for a prepared even state, and similarly for $\epso$. Data are corrected for residual qubit excitations ($1-2~\%$).  
Error bars are smaller than the dot size. Model curves are obtained from $5\,000$ quantum trajectories for each initial state and $\tp$, with quantum efficiencies $\eta=0.25,$ $0.5$, and $1$ for the readout amplification chain. No single value of $\eta$ matches the dependence of $\Fp$ on $\tp$. We attribute this discrepancy to low-frequency fluctuations in the parametric amplifier bias point, not included in the model. 
 \textbf{c,} Concurrence $\conc$ of the two-qubit entangled state obtained by postselection on $\Mp=-1$ (orange) and on $\Mp=+1$ (green squares). Empty symbols correspond to the threshold $\Vthresh$ that maximizes $\Fp$, binning  $\psuccess \sim 50~\%$ of the data into each case. Solid symbols correspond to a threshold $\Vthreshm (\Vthreshp)$ for postselection on $\Mp=-1(+1)$, at which $\epso (\epse) = 0.01$. Concurrence is optimized at $\tp\sim300~\ns$, where  $\psuccess \sim 20\%$ in each case.
We employ maximum-likelihood estimation~\cite{Filipp09} (MLE) to ensure physical density matrices, but concurrence values obtained with and without MLE differ by less than $3\%$  over the full data set.
 \textbf{d, e,} State tomography conditioned on  $\Vp>\Vthreshm$ (\textbf{d}) and   $\Vp<\Vthreshp$ (\textbf{e}), with $\tp=300~\ns$, corresponding to the dark symbols in  \textbf{c}. Figure taken from Ref.~\citenum{Riste13b}.}
\end{figure*}

The ideal continuous parity measurement gradually suppresses the unconditioned density matrix elements $\rho_{ij,kl}=\bra{ij}\rho\ket{kl}$ connecting states with different parity (either $i\neq k$ or $j\neq l$), and leaves all other coherences (off-diagonal terms) and all populations (diagonal terms) unchanged.
The experimental \hl{tomography} reveals the expected suppression of coherence between states of different parity (Figs.~\ref{fig:2_paper5}\textbf{b},\textbf{c}). The temporal evolution of $|\rhoq_{11,10}|$, with near full suppression by $\tp=400~\ns$, is quantitatively matched by a master-equation simulation of the two-qubit system. Tomography also unveils a  non-ideality: albeit more gradually, our parity measurement partially suppresses the absolute coherence between equal-parity states, $|\rhoq_{01,10}|$ and $|\rhoq_{00,11}|$. The effect is also quantitatively captured by the model. Although intrinsic qubit decoherence contributes, the dominant mechanism is the different \hl{AC-Stark phase shift} induced by intra-cavity photons on basis states of the same parity~\cite{Lalumiere10,Tornberg10,Murch13}. This phase shift has both deterministic and stochastic components, and the latter suppresses absolute coherence under ensemble averaging.
We emphasize that this imperfection is technical rather than fundamental. It can be mitigated in the odd subspace by perfecting the matching of $\ChiB$ to $\ChiA$, and in the even subspace by increasing $\chi_\mathrm{A,B}/\kappa$ ($\sim 1.3$ in this experiment).

\subsection{Probabilistic entanglement by measurement and postselection}
The ability to discern  parity subspaces while  preserving coherence within each opens the door to generating entanglement  by parity measurement on $\kpsii$. For every run of the sequence in Fig.~\ref{fig:2_paper5}, we discriminate $\Vint$ using the threshold $\Vthresh$ that maximizes the parity measurement fidelity $\Fp$ (Fig.~\ref{fig:3_paper5}\textbf{a}). Assigning $\Mp=+1\,(-1)$ to $\Vint$ below (above) $\Vthresh$, we bisect the tomographic measurements into two groups, and obtain the density matrix for each.  We quantify the entanglement achieved in each case using concurrence $\conc$ as the metric~\cite{Horodecki09}, which ranges from $0\%$ for an unentangled state to $100\%$ for a Bell state. 
As $\tp$ grows (Fig.~\ref{fig:3_paper5}\textbf{b}), the optimal balance between increasing $\Fp$ at the cost of \hl{measurement-induced dephasing} and intrinsic decoherence is reached at $\sim300~\ns$ (Fig.~\ref{fig:3_paper5}\textbf{c}). \hl{Postselection} on $\Mp=\pm1$ achieves $\conc_{|\Mp=-1} =45\pm 3\%$ and $\conc_{|\Mp=+1}=17\pm 3\%$, with each case occurring with  probability $\psuccess\sim50\%$. The higher performance for $\Mp=-1$ results from lower measurement-induced dephasing in the odd subspace, consistent with Fig.~\ref{fig:2_paper5}.

The entanglement achieved by this probabilistic protocol can be increased with more stringent postselection. Setting a higher threshold
$\Vthreshm$ 
achieves $\conc_{|\Mp=-1} = 77\pm2\%$ but keeps $\psuccess\sim20\%$ of runs.
Analogously, using $\Vthreshp$ achieves $\conc_{|\Mp=+1}=29\pm4\%$ with similar $\psuccess$ (Figs.~\ref{fig:3_paper5}\textbf{d}, \textbf{e}). However, increasing $\conc$ at the expense of reduced $\psuccess$ is not evidently beneficial for QIP. For the many tasks calling for maximally-entangled qubit pairs (ebits), one may use an optimized distillation protocol~\cite{Horodecki09} to prepare one ebit from $N=1/\LN(\rho)$ pairs in a partially-entangled state $\rho$,
where $\LN$ is the \hl{logarithmic negativity}~\cite{Horodecki09}. The \hl{efficiency} $\Ratee$ of ebit generation would be $\Ratee= \psuccess   \LN\left(\rhoq\right)$. 
For postselection on $\Mp=-1$, we calculate $\Ratee=0.31~\ebiteff$ using $\Vthresh$ and $\Ratee=0.20~\ebiteff$ using $\Vthreshm$. Evidently, increasing entanglement at the expense of reducing $\psuccess$ is counterproductive in this context.

\begin{figure*}
\centering
\includegraphics[width=\columnwidth]{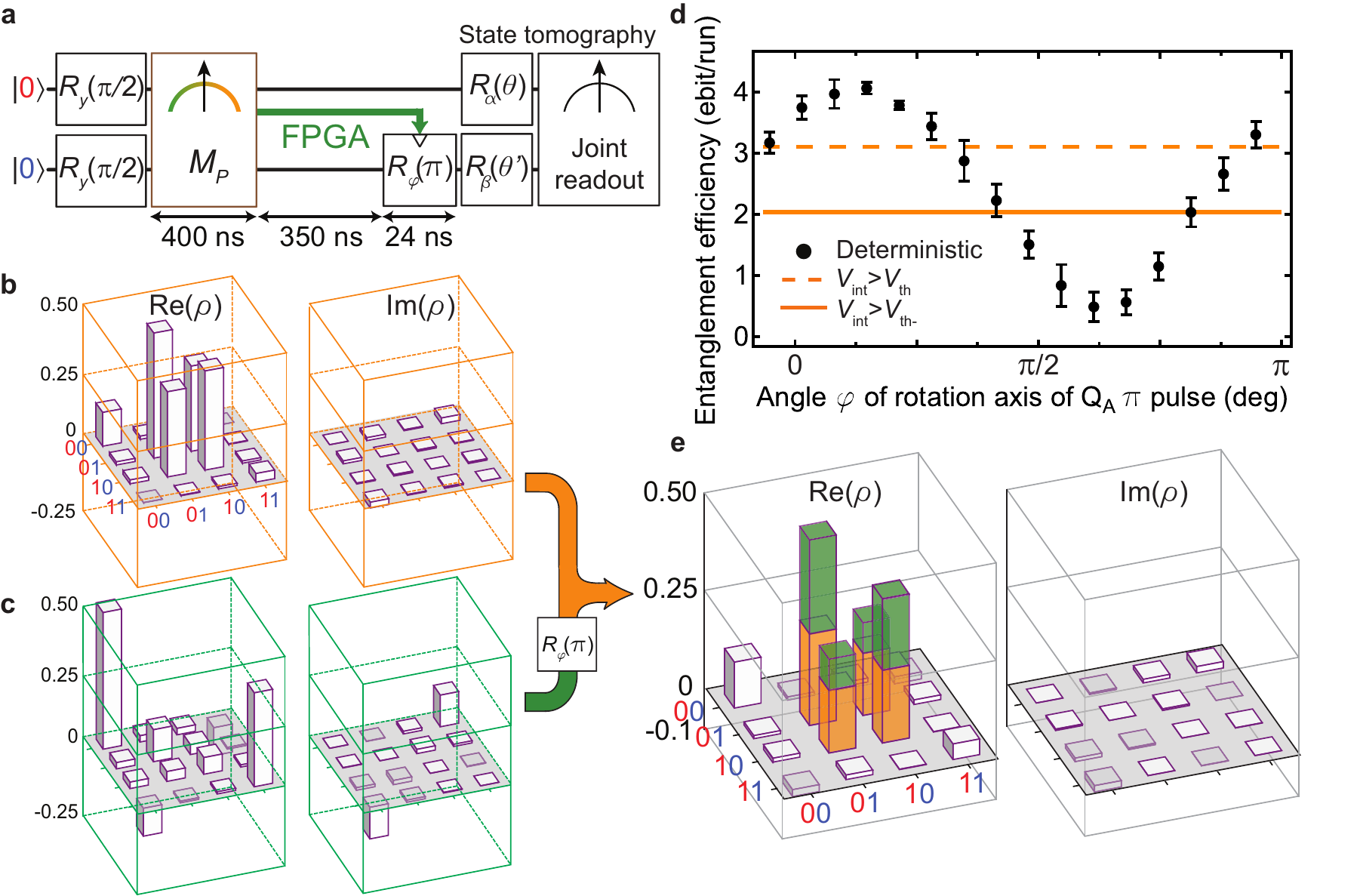}
\caption{\label{fig:4_paper5} \textbf{Deterministic entanglement generation using feedback.}  \textbf{a,} 
We close a digital feedback loop by triggering (via the FPGA) a $\pi$ pulse on $\QA$ conditional on parity measurement result $\Mp=+1$. This $\pi$ pulse switches the two-qubit parity from even to odd, and allows the deterministic targeting of $\OddBellp=(\ket{01}+\ket{10})/\sqrt{2}$.
 \textbf{b, c,} Parity measurement results $\Mp=-1$ and $\Mp=+1$ each occur with $\sim50\%$ probability. The deterministic AC Stark phase acquired between $\ket{01}$ and $\ket{10}$ during parity measurement (due to residual mismatch between $\ChiA$ and $\ChiB$) is compensated by a global phase rotation in  the tomography pulses. A different AC Stark phase is acquired between $\ket{00}$ and $\ket{11}$, resulting in the state shown in  \textbf{c},  with the maximal overlap with even Bell state $[\ket{00}+\exp(-  i \varphi_\mathrm{e})\ket{11}]/\sqrt{2}$ at $\varphi_\mathrm{e}=0.73\pi$.
 \textbf{d,} Generation rate of entanglement using feedback, as a function of the phase $\varphi$ of the $\pi$ pulse. The deterministic entanglement generation efficiency outperforms the efficiencies obtained with  postselection (Fig.~\ref{fig:3_paper5}). Error bars are the standard deviation of $7$ repetitions of the experiment at each $\varphi$.  \textbf{e,} Full state tomography for deterministic entanglement [$\varphi=(\pi-\varphi_\mathrm{e})/2$], achieving fidelity $\BraOddBellp \rho \OddBellp = 66\%$ to the targeted $\OddBellp$, and concurrence $\conc = 34\%$. Colored bars highlight the contribution from cases $\Mp=-1$ (orange) and $\Mp=+1$ (green). Figure taken from Ref.~\citenum{Riste13b}.}
\end{figure*}

\subsection{Deterministic entanglement by measurement and feedback}
Motivated by the above observation, we finally demonstrate the use of feedback control to transform entanglement by parity measurement from probabilistic to deterministic, i.e., $\psuccess=100\%$. While initial proposals in cQED focused on analog feedback schemes~\cite{Sarovar05}, here we adopt a digital strategy. Specifically, we use our homebuilt programmable controller (section~\ref{sec:fbcontrollers} to apply a $\pi$ pulse on $\QA$ conditional on measuring $\Mp=+1$ (using $\Vthresh$, Fig.~\ref{fig:4_paper5}). In addition to switching the two-qubit parity, this pulse lets us choose which odd-parity Bell state to target by selecting the phase $\varphi$ of the conditional pulse. To optimize \hl{deterministic entanglement}, we need to maximize overlap to the same odd-parity Bell state for $\Mp=-1$ (Fig.~\ref{fig:4_paper5}\textbf{b}) as for $\Mp=+1$ (Fig.~\ref{fig:4_paper5}\textbf{c}). For the targeted state $\OddBellp$, this requires cancelling the deterministic AC Stark phase $\varphi_\mathrm{e}=0.73\pi$ accrued between $\ket{00}$ and $\ket{11}$ when $\Mp=+1$. This is accomplished by choosing $\varphi=(\pi-\varphi_\mathrm{e})/2$, which clearly maximizes the entanglement obtained when no postselection on $\Mp$ is applied (Figs.~\ref{fig:4_paper5}\textbf{c}, \textbf{d}). The highest deterministic $\conc=34\%$ achieved is lower than for our best probabilistic scheme, but the boost to $\psuccess=100\%$ achieves a higher $\rateE=0.41~\ebiteff$. 

A parallel development realized the probabilistic entanglement by measurement between two qubits in separate 3D cavities~\cite{Roch14}, establishing the first quantum connection between remote superconducting qubits. In another two-qubit, single-cavity system, feedback has been recently applied to enhance the fidelity of the generated entanglement~\cite{Liu15}. 
Following the first realizations in 3D cQED, parity measurements have been implemented using an ancillary qubit~\cite{Saira14, Chow14} in 2D. Compared to the cavity-based scheme, the use of an ancilla evades measurement-induced dephasing and is better suited to scaling to larger circuits.

\section{Conclusion}
We have presented the first implementation of digital feedback control in superconducting circuits, and its evolution to faster, simpler, and more configurable feedback loops. In particular, we showed the use of digital feedback for fast and deterministic qubit reset and for deterministic generation of entanglement by parity measurement. Considering the vast range of applications for feedback in quantum computing, we hope that this development is just the start of an exciting new phase of measurement-assisted digital control in solid-state quantum information processing. 

\section*{Acknowledgments}
We thank all the collaborators who have contributed to the experiments here presented:  J.~G.~van Leeuwen, C.~C.~Bultink, M.~Dukalski, C.~A.~Watson, G.~de Lange, H.-S.~Ku, M.~J.~Tiggelman,  K.~W.~Lehnert, Ya.~M.~Blanter, and R.~N.~Schouten. We acknowledge L.~Tornberg and G.~Johansson for useful discussions. Funding for this research was provided by the Dutch Organization for Fundamental Research on Matter (FOM), the Netherlands Organization for Scientific Research (NWO, VIDI scheme), and the EU FP7 projects SOLID and SCALEQIT.

\end{document}